\documentclass[showpacs,prb,aps,twocolumn]{revtex4-1}

\usepackage{amsmath,amsfonts,bm,graphicx,verbatim,mathrsfs,units,psfrag}
\usepackage[colorlinks=true,citecolor=blue]{hyperref}

\newcommand{\rhohat}{\hat{\rho}}

\newcommand{\hatH}{\hat{H}}

\newcommand{\bra}[1]{\langle #1|}
\newcommand{\ket}[1]{|#1\rangle}
\newcommand{\hc}{\textrm{h.c.}}

\newcommand{\ew}[1]{\langle #1\rangle}
\newcommand{\of}[1]{\left(#1\right)}
\renewcommand{\epsilon}{\varepsilon}
\newcommand{\eps}{\varepsilon}
\newcommand{\tb}{\bar{t}}

\begin{document}
\title{Waiting time distributions of noninteracting fermions on a tight-binding chain}

\author{Konrad H.\ Thomas}
\author{Christian Flindt}
\affiliation{D\'epartement de Physique Th\'eorique, Universit\'e de Gen\`eve, 1211 Gen\`eve, Switzerland}
\date{\today}

\begin{abstract}
We consider the distribution of waiting times between non-interacting fermions on a tight-binding chain. We calculate the waiting time distribution for a quantum point contact and find a cross-over from Wigner-Dyson statistics at full transmission to Poisson statistics close to pinch-off as predicted by scattering theory. In addition, we consider several quantum dot structures for which we can associate oscillations in the waiting time distributions to internal energy scales of the scatterers. A detailed comparison with scattering theory and generalized master equations is provided. We focus on mesoscopic conductors, but our tight-binding models may also be realized in cold atomic gases.
\end{abstract}



\maketitle

\section{Introduction}
\label{sec:Introduction}

Quantum electronics is a rapidly developing field of research.\cite{Gabelli2006,Feve2007,Bocquillon2012,Bocquillon2013,Fletcher2013,Dubois2013,Gaury2014} Experimental progress is currently being made towards the detection of single electrons coherently traversing a mesoscopic structure. With these advances, it may soon be possible not only to measure mean currents, shot noise,\cite{Blanter2000} and the first few higher-order current correlation functions,\cite{Reulet2003,Bomze2005,Gershon2007,Timofeev2007} but even the full statistical distribution of transferred charge in a quantum-coherent conductor may become accessible. With this in mind, it is now the right moment to develop statistical tools to analyze and interpret such experimental data.
 
In one approach, the full counting statistics (FCS) of transferred charge is investigated.\cite{Levitov1993,Levitov1996,Nazarov2003} Traditionally, the charge fluctuations are integrated over a long period of time and the zero-frequency current cumulants are measured. However, electrical fluctuations at finite times have become of increasing interest and the FCS of charge transfer in quantum dots has now been measured in several single-electron counting experiments.\cite{Gustavsson2006,Sukhorukov2007,Gustavsson2009,Flindt2009,Fricke2010a,Fricke2010b,Ubbelohde2012,Maisi2014,Hassler2008} In addition, to characterize short-time fluctuations the distribution of waiting times between transferred charges has recently been proposed as an alternative to FCS.\cite{Brandes2008,Welack2009,Albert2011,Albert2012,Thomas2013,Rajabi2013,Dasenbrook2013,Albert2014} Waiting time distributions (WTDs) are known in quantum optics,\cite{Vyas1988,Carmichael1989} but are now also being used in quantum transport. Methods have become available for evaluating the WTDs of electronic systems described by generalized master equations\cite{Brandes2008,Welack2009,Albert2011,Thomas2013,Rajabi2013} (GMEs)~or~by~scattering~theory.\cite{Albert2012,Dasenbrook2013}

In this work we consider the WTD of non-interacting fermions on a finite-size tight-binding chain with a fixed number of particles. A central task is to understand if such a system can mimic the quantum transport in a conductor that is coupled to large particle reservoirs. We develop a method to evaluate the WTD based on work by Sch\"onhammer who considered the FCS of non-interacting fermions in a one-dimensional tight-binding system.\cite{Schonhammer2007,Inhester2009,Schonhammer2009} Similar tight-binding approaches have been used to evaluate the FCS in disordered free-fermion systems\cite{Levine2012} as well as the finite-frequency noise\cite{Jonckheere2012} of driven single-electron emitters.\cite{Mahe2010,Albert2010,Parmentier2012}

In the following, we begin by occupying states in the left part of the tight-binding chain to establish a flow of particles via the central scatterer to the right side of the chain. After a transient behavior, a quasi-stationary regime is reached during which the particle current through the scatterer is constant and we can compute the WTD. We evaluate the WTD for a quantum point contact (QPC) as well as a number of different quantum dot structures. We compare our results with predictions based on scattering theory and find excellent agreement.\cite{Albert2012} This is an important check of our method, which may thus serve as a stepping stone towards a theory of WTDs for interacting fermions, for example based on density matrix renormalization group (DMRG) techniques.\cite{Daley2004,Bohr2006,Carr2011,Chien2012a,Chien2013} Under appropriate conditions, we also find good agreement with calculations using generalized master equations (GME).\cite{Brandes2008,Thomas2013} We focus here on electronic conductors, but our tight-binding models may also be realized in cold atomic gases.\cite{Brantut2012,Krinner2014}

Our paper is organized as follows: In Sec.~\ref{sec:Formalism} we introduce the basic theory of WTDs. We describe our tight-binding model as well as our method for calculating WTDs. In Sec.~\ref{sec:Results} we illustrate the method with several applications.  We first calculate the WTD for a QPC and find a cross-over from Wigner-Dyson statistics at full transmission to Poisson statistics close to pinch-off as predicted by scattering theory.\cite{Albert2012}  Next, we consider several quantum dot structures. We calculate the WTD for a single as well as a double quantum dot, which may enclose a magnetic flux if the quantum dots are arranged in parallel. We show how oscillations in the WTD may be associated with internal energy scales of the scatterer. For the quantum dot systems we find good agreement with GME calculations.\cite{Brandes2008,Thomas2013} Finally, we consider the WTD for a bipartite chain, where a gap opens in the transmission spectrum as the chain becomes long. Our work is summarized in Sec.~\ref{sec:Conclusion}. Several technical details of our calculations are described in the appendices (\ref{App:A}-\ref{App:D}).

\begin{figure*}
\includegraphics[width=0.95\textwidth]{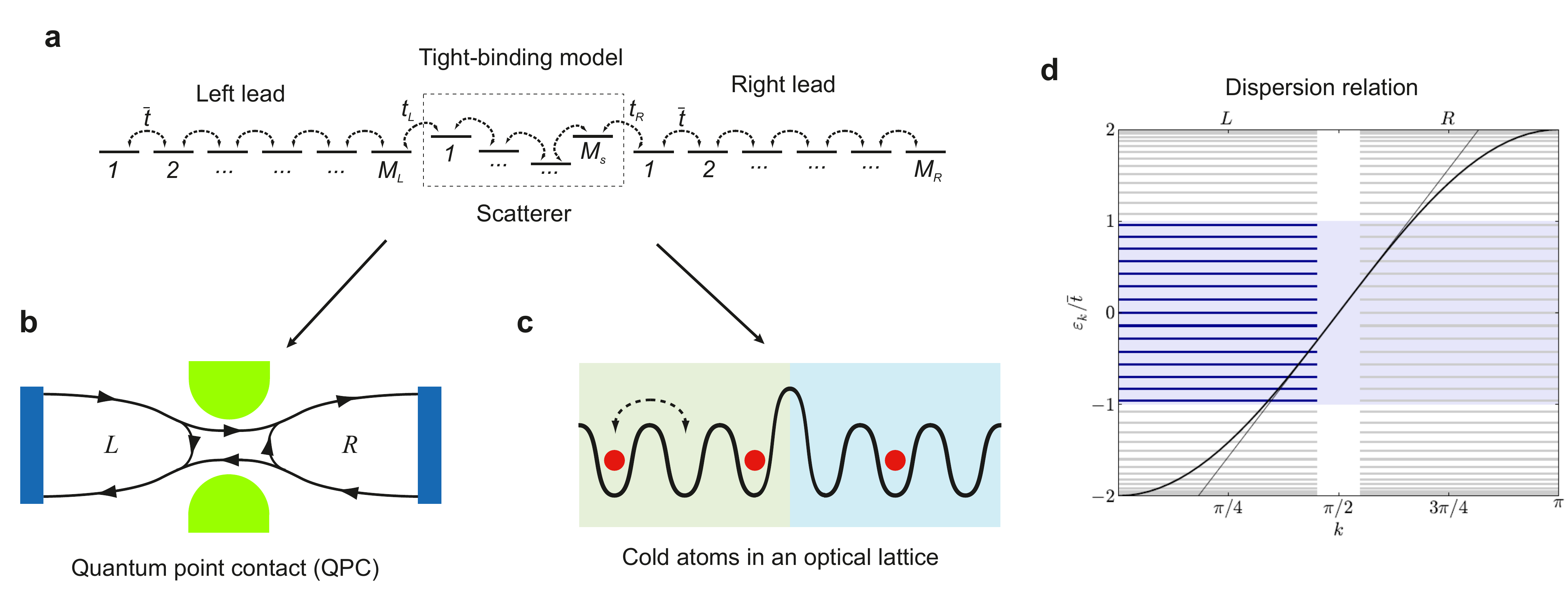}
\caption{(Color online) Tight-binding model, physical realizations, and dispersion relation. {\bf a}, Left and right leads connected to a central scatterer. The leads consist of tight-binding chains with $M_{\alpha}$ sites, $\alpha=L,R$, and nearest-neighbor hopping amplitude $\tb$. Each lead is coupled with hopping amplitude $t_\alpha$ to the scatterer consisting of $M_s$ sites. {\bf b}, The tight-binding model may represent a mesoscopic conductor coupled to electronic leads, for instance a QPC as shown here. {\bf c}, Cold atoms in an optical lattice are another possible realization of the tight-binding model. {\bf d}, Cosine-dispersion $\eps_k$ of the leads (thick line), see Eq.~(\ref{eq:cos_disp}). The dispersion relation is linear (thin line) around $k=\pi/2$. The horizontal lines are the eigenenergies of the left ($L$) and right ($R$) leads. The states of the left lead marked with blue (in the shaded region) are initially occupied.}
\label{Fig:TBsys}
\end{figure*}

\section{Waiting Time Distributions}
\label{sec:Formalism}

We consider the generic tight-binding model depicted in Fig.~\ref{Fig:TBsys}a. It consists of left and right tight-binding leads connected to a central scatterer. We consider situations where non-interacting fermions are transferred from the left to the right lead via the scatterer. We are interested in the distribution ${\cal W}(\tau)$ of waiting times $\tau$ between subsequent particles appearing at a given site in the right lead once a stationary particle flow through the scatterer has been established. Physically, our model may represent a mesoscopic conductor coupled to electronic leads, for example the QPC in Fig.~\ref{Fig:TBsys}b, which we describe as a weak tunneling amplitude between the two leads. Other realizations include cold fermionic atoms in an optical lattice as depicted in Fig.~\ref{Fig:TBsys}c.

A central quantity in our work is the idle time probability (ITP) $\Pi(t_0,\tau)$, i.~e.~the probability of detecting no transferred particles during the time interval $[t_0, t_0+\tau]$.\cite{Vyas1988,Albert2012} In general, the ITP depends both on $t_0$ and $\tau$. However, for stationary processes the ITP is a function only of the length $\tau$ of the time interval, such that $\Pi(t_0,\tau)=\Pi(\tau)$. In this case, the WTD can be expressed as\cite{Vyas1988,Albert2012}
\begin{equation}
{\cal W}(\tau)=\ew{\tau}\frac{d^2}{d\tau^2}\Pi(\tau),
\label{eq:WTD}
\end{equation}
where
\begin{equation}
\ew{\tau}=\int_0^\infty d\tau {\cal W}(\tau)\tau=-\frac{1}{\dot{\Pi}(0)}
 \label{eq:meanWT}
\end{equation}
is the mean waiting time. In a conventional quantum transport setup consisting of a mesoscopic conductor connected to (infinitely large) external electronic reservoirs, the transport is stationary in the absence of any explicit time-dependence and the relation above holds. In contrast, we consider here a situation in which the external particle reservoirs are not infinitely large and the number of particles in the isolated system is fixed. However, as we will see, after a flow of particles through the scatterer has been established, a period of time exists where the transport is quasi-stationary and $\Pi(t_0,\tau)\simeq \Pi(\tau)$ is independent of $t_0$. Under these conditions, we may evaluate the WTD using Eq.~(\ref{eq:WTD}).

\subsection{Tight-binding Hamiltonian}

The tight-binding Hamiltonians that we consider could originate from a problem of non-interacting spinless fermions governed by the single-particle Hamiltonian
\begin{equation}
\hatH = -\frac{\hbar^2}{2m_e}\frac{d^2}{d x^2}+V(x).
\label{eq:sing_ham}
\end{equation}
Here, the potential $V(x)$ is only non-zero inside the scattering region. This problem can be discretized on a lattice with lattice spacing $a$.\footnote{The continuum description can be recovered by taking the limits $a\to0,\; M\to\infty$, where $M$ is the number of lattice sites, keeping the length of the system $L=Ma$ constant.} The single-particle wave-function $\psi(x)$ takes the value
\begin{equation}
\psi_l=\psi(x=x_l)
\end{equation}
on lattice site number~$l$ with $x_l=la$. Similarly, for the potential $V(x)$ we define
\begin{equation}
V_l=V(x=x_l).
\end{equation}
Finally, for the kinetic part of the Hamiltonian we use the standard discretization\cite{Datta1995}
\begin{equation}
-\frac{\hbar^2}{2m_e}\frac{d^2}{d x^2}\psi(x)|_{x=x_l}\simeq -\frac{\hbar^2}{2m_ea^2}\left(\psi_{l+1}-2\psi_l+\psi_{l-1}\right)
\label{eq:kin-dis}
\end{equation}
from which we can identify
\begin{equation}
\tb=\frac{\hbar^2}{2m_ea^2}
\end{equation}
as the tunneling amplitude between neighboring sites. The discretization in Eq.~(\ref{eq:kin-dis}) introduces the constant on-site energy $2\tb$, which is absorbed into the potential by redefining it as $V_l+2\tb\rightarrow V_l$. We then arrive at a tight-binding Hamiltonian of the form
\begin{equation}
\hatH_{\mathrm{tb}}=\hatH_{\mathrm{scat}}+\hatH_{\mathrm{leads}}+\hatH_{\mathrm{tun}}.
\label{eq:H_tb}
\end{equation}
The Hamiltonian of the scatterer reads
\begin{equation}
\hatH_{\mathrm{scat}}=-\tb\sum_{m=1}^{M_s-1} \left({\ket{m}\bra{m+1}+\hc}\right)+\sum_{m=1}^{M_s} V_m\ket{m}\bra{m},
\label{eq:H_scat}
\end{equation}
having assumed that it consists of $M_s$ sites, labeled as $\{\ket{m}\}$. Specific expressions for the scatterer are given in the examples below. The Hamiltonian of the leads
\begin{equation}
\hatH_{\mathrm{leads}}=\sum_{\alpha=L,R}\hatH_{\alpha}
\label{eq:H_leads}
\end{equation}
consists of the two parts
\begin{equation}
\hatH_{\alpha}=-\tb\sum_{m=1}^{M_\alpha-1}\ket{m,\alpha}\bra{m+1,\alpha}+\hc,
\label{eq:H_alpha}
\end{equation}
where lead $\alpha=L,R$ contains $M_\alpha$ sites labeled as $\{\ket{m,\alpha}\}$. Finally, if not otherwise stated, tunneling between the scatterer and the leads is described by the Hamiltonian
\begin{equation}
\hatH_{\mathrm{tun}}=-t_L\ket{M_L,L}\bra{1}-t_R\ket{1,R}\bra{M_s}+\hc,
\label{eq:H_tunnel}
\end{equation}
which connects the rightmost (leftmost) site of the left (right) lead to the leftmost (rightmost) site of the scatterer with tunneling amplitude $t_L$ ($t_R$). Below, we will in general allow $t_{L(R)}$ to be different from $\tb$.

\subsection{Idle time probability}
\label{subsec:ITP}

We are now ready to calculate the ITP. In what follows, we start out by initializing the left lead with a fixed number of particles. We consider the system at zero temperature, although a finite temperature can be implemented following Ref.~\onlinecite{Schonhammer2009}. At time $t=0$, we connect the scatterer to the leads as described by the tunneling Hamiltonian in Eq.~(\ref{eq:H_tunnel}). Having established the connection, particles start to flow from the left lead to the right lead via the scatterer. After a transient behavior, the flow of particles becomes quasi-stationary for a period of time, limited by the finite number of particles and the finite size of the leads. Still, in this quasi-stationary regime, we may calculate the ITP together with the WTD using Eq.~(\ref{eq:WTD}).

To evaluate the ITP we first analyze the Hamiltonian of the leads given by Eq.~(\ref{eq:H_alpha}). The eigenstates of each lead Hamiltonian are the standing-wave solutions\cite{Schonhammer2007}
\begin{equation}
\ket{k^j_\alpha}=\sqrt{\frac{2}{M_\alpha+1}}\sum_{m=1}^{M_\alpha}\sin(k_\alpha^j m)\ket{m,\alpha},\hspace{0.2cm}
k_\alpha^j=\frac{j\pi}{M_\alpha+1},
\label{eq:lead_states}
\end{equation}
$j=1,\ldots,M_\alpha$, which vanish outside the leads. The corresponding dispersion relation reads
\begin{equation}
\eps_{k_\alpha^j}=-2\tb\cos(k_\alpha^j),
\label{eq:cos_disp}
\end{equation}
such that the group velocity becomes
\begin{equation}
v_k=\frac{1}{\hbar}\frac{\partial\eps_k}{ \partial k}=\frac{2\tb}{\hbar}\sin(k).
\label{eq:group_velo}
\end{equation}
We take the same number of sites for the two leads
\begin{equation}
M_L=M_R=M.
\end{equation}
Additionally, we occupy $N_{0}$ states of the left lead in the linear part of the dispersion relation centered around $k=\pi/2$, see Fig.~\ref{Fig:TBsys}d.  Here the group velocity is approximately constant, since it can be expanded as
\begin{equation}
v_{k}=v_F+\mathcal{O}((k-\pi/2)^2)
\label{eq:group_velo_approx}
\end{equation}
close to $k=\pi/2$, where
\begin{equation}
v_F=\frac{2\tb}{\hbar}
\label{eq:Fermi velo}
\end{equation}
is the Fermi velocity. The constant group velocity is important for our calculations of the ITP. For the rest of the paper, we set $\hbar=1$.

To evaluate the ITP, we consider a particular site of the right lead $\ket{M_d,R}$ and calculate the probability of detecting no particles at this site during the \emph{temporal} interval $[t_0, t_0+\tau]$. In the quasi-stationary regime, the ITP depends only on the length of the time interval $\tau$ and is independent of $t_0$. Importantly, since all particles move with the Fermi velocity $v_F$ from the left lead towards site $\ket{M_d,R}$ in the right lead, we may instead consider the probability of detecting no particles in the \emph{spatial} interval between sites $\ket{\lceil M_d- v_F\tau\rceil,R}$ and $\ket{M_d,R}$, where $\lceil\cdot\rceil$ denotes ceiling (or equivalently, due to the quasi-stationary conditions, between sites $\ket{M_d,R}$, and $\ket{\lfloor M_d+ v_F\tau\rfloor,R}$, where $\lfloor\cdot\rfloor$ denotes flooring).

The probability of finding a particular particle between sites $\ket{1,R}$ and $\ket{\lfloor 1+ v_F\tau\rfloor,R}$, taking $M_d=1$ from now on, is given by the expectation value of the operator\cite{Hassler2008,Albert2012}
\begin{equation}
\hat{\cal Q}_\tau= \sum_{m=1}^{M_R}\ket{m,R}\bra{m,R}\hspace{0.1cm}\Theta(m-v_F\tau).
\label{eq:Q_mat}
\end{equation}
The probability of \emph{not} observing the particle is then given by the (single-particle) expectation value of $1-\hat{\cal Q}_\tau$. Below, we consider $N_0$ particles in the system (rather than a single particle), and the operator $1-\hat{\cal Q}_\tau$ must act on all particles in the many-body state. The ITP therefore becomes
\begin{equation}
\Pi(\tau)=\bra{\Psi_S(t_0)}\bigotimes_{j=1}^{N_0}\left[1-\hat{\cal Q}_\tau(\tau)\right]\ket{\Psi_S(t_0)}.
\label{eq:ITP1}
\end{equation}
Here we have evolved the initial many-body state $\ket{\Psi_S(0)}$ at $t=0$ with all particles in the left lead to a time $t_0$, where the transport has become quasi-stationary. Additionally, we have defined
\begin{equation}
\hat{\cal Q}_\tau(\tau) =  e^{i\hatH_{\mathrm{tb}} \tau}\hat{\cal Q}_\tau e^{-i\hatH_{\mathrm{tb}} \tau}.
\end{equation}

Importantly, as we are dealing with non-interacting fermions, the many-body state $\ket{\Psi_S(t_0)}$ is a Slater determinant (as indicated with the subscript $S$), which at $t=0$ is constructed from the filled single-particle states of the left lead. The expectation value of a product of single-particle operators with respect to a Slater determinant can itself be written as a determinant\cite{Hassler2008,Albert2012} and Eq.~(\ref{eq:ITP1}) thereby simplifies to
\begin{equation}
\Pi(\tau)=\det[\mathbf{1}-\mathbf{Q}_\tau(\tau)].
\label{eq:ITP2}
\end{equation}
The matrix elements of $\mathbf{Q}_\tau(\tau)$ are taken with respect to the initially filled states of the left lead,
\begin{equation}
[\mathbf{Q}_\tau(\tau)]_{k_L^m,k_L^n}=\ew{k_L^m( t_0)|\hat{\cal Q}_\tau(\tau) |k_L^n (t_0)},
\label{eq:Qelem}
\end{equation}
which have been evolved from $t=0$ to $t_0$, i.~e.
\begin{equation}
\ket{k_L^n (t_0)}= e^{-i\hatH_{\mathrm{tb}} t_0}\ket{k_L^n}.
\end{equation}

With these expressions at hand, we are now in position to state our final result for the WTD. To this end, we recall Jacobi's formula for the derivative of a determinant
\begin{equation}
\frac{d}{dt} \det[\mathbf{A}(t)]=\textrm{Tr}[\textrm{adj}\{\mathbf{A}(t)\}\dot{\mathbf{A}}(t)],
\end{equation}
where $\textrm{adj}\{\mathbf{A}\}$ is the adjugate of $\mathbf{A}$ and $\dot{\mathbf{A}}(t)=\frac{d}{dt}\mathbf{A}(t)$. For invertible matrices, the adjugate reads \begin{equation}
\textrm{adj}\{\mathbf{A}\}=\det[\mathbf{A}]\mathbf{A}^{-1}.
\end{equation}
For the mean waiting time, we then find
\begin{equation}
\langle\tau\rangle=\frac{1}{\textrm{Tr}[\dot{\mathbf{Q}}_0(0)]},
\label{eq:meanWTfinal}
\end{equation}
having used Eq.~(\ref{eq:meanWT}). From Eq.~(\ref{eq:WTD}), we moreover find
\begin{equation}
{\cal W}(\tau) = \Pi(\tau) \frac{\mathrm{Tr}^2[\mathbf G\dot{\mathbf{Q}}_\tau(\tau)]-\mathrm{Tr}[\{\mathbf G\dot{\mathbf{Q}}_\tau(\tau)\}^2+\mathbf G\ddot{\mathbf{Q}}_\tau(\tau)]}{\textrm{Tr}[\dot{\mathbf{Q}}_0(0)]},
\label{eq:WTDfinal}
\end{equation}
where we have defined
\begin{equation}
\mathbf G = [\mathbf{1}-\mathbf{Q}_\tau(\tau)]^{-1}.
\end{equation}
In addition, we have
\begin{equation}
\dot{\mathbf{Q}}_\tau(\tau)=i[\mathbf{H}_{\mathrm{tb}},\mathbf{ Q}_\tau(\tau)]+e^{i\mathbf{H}_{\mathrm{tb}} \tau}(\partial_\tau\mathbf{Q}_\tau) e^{-i\mathbf{H}_{\mathrm{tb}} \tau},
\end{equation}
where $\mathbf{H}_{\mathrm{tb}}$ is the matrix representation of $\hatH_{\mathrm{tb}}$. Using Eq.~(\ref{eq:WTDfinal}) we may calculate the WTD for an arbitrary scatterer connecting the left and right leads.

\section{Results}
\label{sec:Results}

In this section we illustrate our method by calculating the WTDs for a number of scatterers. First, we consider a QPC. Special attention is paid to the time it takes the system to reach the quasi-stationary regime. We discuss several technical details related to our calculations. In the following examples we consider a single-level quantum dot as well as a double quantum dot. The two quantum dots can be arranged with the levels either in series or in parallel, such that a magnetic flux can be enclosed. We also consider a bipartite chain, where a gap opens in the transmission spectrum as the chain becomes long. We compare our numerical results to methods based on scattering theory\cite{Albert2012} or generalized master equations.\cite{Brandes2008,Thomas2013}

\subsection{Quantum point contact}
\label{subsec:QPC}

We first consider a QPC which directly couples the left and right leads by the tunneling amplitude $t_\textrm{QPC}$. The Hamiltonian $\hatH_\textrm{scat}$ in Eq.~(\ref{eq:H_scat}) is then absent and the tunneling Hamiltonian in Eq.~(\ref{eq:H_tunnel}) is simply
\begin{equation}
\hatH_\textrm{tun}=-t_\textrm{QPC}\ket{M_L, L}\bra{1,R}+\hc
\label{eq:H_QPC}
\end{equation}
To begin with the two leads are unconnected and we prepare the left lead with $N_0=M/3$ particles in the linear region of the dispersion relation (recalling that $M$ is the number of sites in each lead). Specifically, we fill the states with energies in the interval $[-V/2,V/2]$, where $V=2\tb$. This value of $V$ is chosen as a trade-off between, on the one hand, staying within the linear region of the dispersion relation and, on the other hand, having a large energy window which reduces the computation time.

At $t=0$, the two leads are connected and particles begin to flow from the left lead to the right lead. The number of particles in the right lead can be expressed as
\begin{equation}
N_R(t)=\sum_{k_L^j\,\mathrm{occup.}}\bra{k_L^j}\hat{P}_R(t)\ket{k_L^j}, t\geq 0,
\label{eq:N_R}
\end{equation}
where the sum runs over the initially occupied states of the left lead and we have introduced the projection operator onto the right lead
\begin{equation}
\hat{P}_R(t)=e^{i\hatH_{\mathrm{tb}}t}\left[\sum_{m=1}^{M_R}\ket{m,R}\bra{m,R}\right]e^{-i\hatH_{\mathrm{tb}}t}.
\label{eq:P_R}
\end{equation}
The time-dependent particle current running into the right lead is then
\begin{equation}
I_R(t)=\frac{d}{dt}N_R(t)=i\!\!\sum_{k_L^j\,\mathrm{occup.}}\bra{k_L^j}[\hatH_{\mathrm{tb}},\hat{P}_R(t)]\ket{k_L^j},
\label{eq:I_R}
\end{equation}
which can be evaluated in a straightforward manner.

\begin{figure}
	\includegraphics[width=0.95\columnwidth]{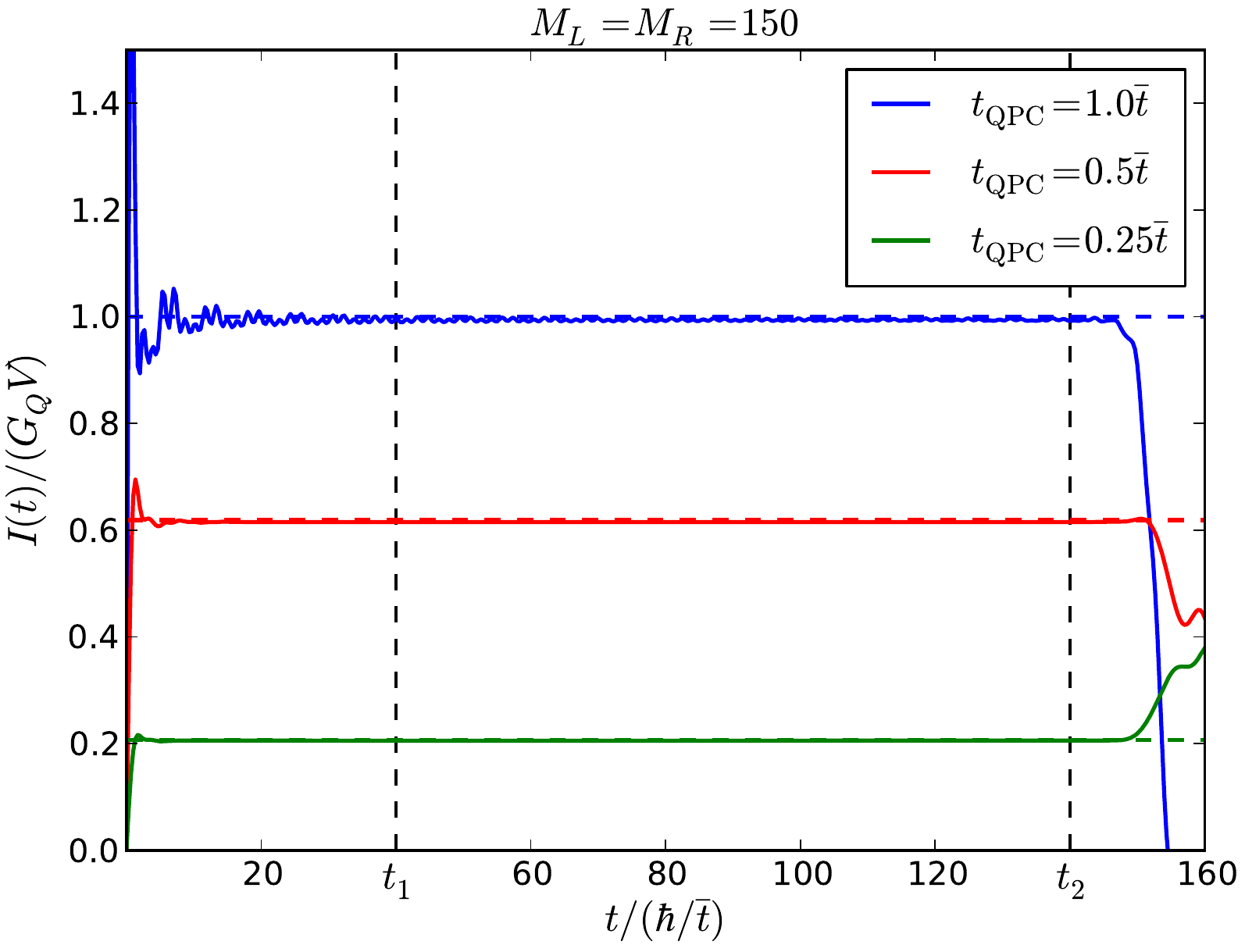}
	\caption{\label{Fig:I_QPC} (Color online) Time-dependent current through the QPC. The QPC is opened at $t=0$. Numerical results (solid lines) are shown for three different tunneling amplitudes~$t_\textrm{QPC}$. The horizonal dashed lines show the expected stationary current based on scattering theory. The vertical dashed lines mark the ``window of opportunity'', $[t_1,t_2]$, during which the transport is quasi-stationary and we can evaluate the WTD. Initially, the left lead is occupied by $N_0=50$ particles.}
\end{figure}

In Fig.~\ref{Fig:I_QPC} we show the time-dependent particle current for three different values of the QPC tunneling amplitude. After the connection is established at $t=0$, the current goes through a transient behavior before reaching a quasi-stationary value around the time $t_1$ (marked with a vertical dashed line), see also Refs.~\onlinecite{Schonhammer2007,Chien2012b,Chien2014}. The value of the quasi-stationary current is consistent with predictions based on scattering theory (horizontal dashed lines) as we discuss below. The current stays constant until the time $t_2$, when finite-size effects become visible. However, the quasi-stationary regime $[t_1, t_2]$ provides us with a ``window of opportunity'' during which we may evaluate the WTD using Eq.~(\ref{eq:WTDfinal}).

In the following, all results are obtained by averaging over at least three calculations with different choices of starting times $t_0\in[t_1,t_2]$ for evaluating the WTD. The WTD is calculated for discrete times $\tau_m=m/v_F$, $0\leq m\leq M$, due to the discretization of the leads. The length of the leads, determining the length of the quasi-stationary regime $[t_1,t_2]$, must be chosen such that the WTD approaches zero before $t_2$ is reached. Typical values are $t_1\approx M/3v_F$ and $t_2\lesssim M/v_F$.

\begin{figure}
	\includegraphics[width=0.95\columnwidth]{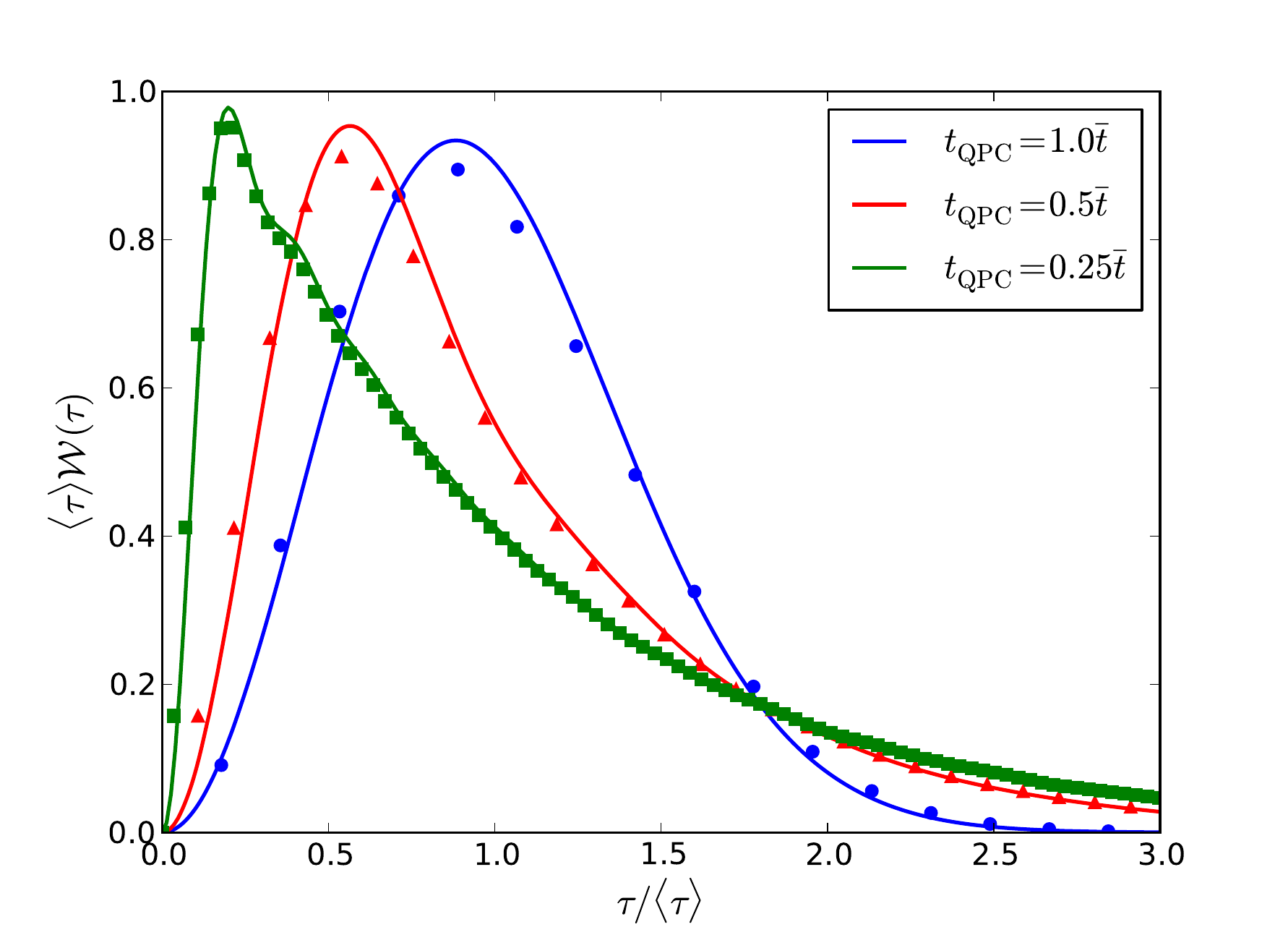}
	\caption{\label{Fig:Res_QPC} (Color online) WTD for the QPC. Numerical results (symbols) for three different tunneling amplitudes $t_\textrm{QPC}$. For all three curves, the waiting time is given in units of the mean waiting time $\ew{\tau}$. The numerical results agree well with predictions based on scattering theory (solid lines), showing a cross-over from Wigner-Dyson statistics at full transmission (blue curve) to Poissonian statistics with an approximately exponential WTD (green curve) at low transmissions.}
\end{figure}

In Fig.~\ref{Fig:Res_QPC} we show WTDs for the three different values of the tunneling amplitude $t_\textrm{QPC}$. We have rescaled the horizontal axis by the mean waiting time $\ew{\tau}$ such that the mean waiting time of each (rescaled) distribution is unity. In the linear part of the dispersion relation, the size of the energy window $V$ only determines the mean waiting time $\bar{\tau}=h/V$ of the particles in the incoming many-particle state as shown in Ref.~\onlinecite{Albert2012}. We have checked that the results in Fig.~\ref{Fig:Res_QPC} do not depend on the value of~$V$. The suppression of the WTDs at short times reflects the fermionic statistics of the particles, which prevents two particles from being detected at the same time.

The calculations in Ref.~\onlinecite{Albert2012} are based on scattering theory with semi-infinite leads connected to the scatterer. One important prediction is that the WTD for a QPC should exhibit a cross-over from a Wigner-Dyson distribution at full transmission to Poisson statistics close to pinch-off. This prediction is confirmed by our numerical results. Remarkably, at low transmissions the tight-binding results reproduce the small oscillatory features in the WTD with period $\bar{\tau}$, also found in Ref.~\onlinecite{Albert2012}.

To make the comparison with scattering theory quantitative, we calculate the scattering amplitude of our setup and evaluate the WTD using the method developed in Ref.~\onlinecite{Albert2012} (see also App.~\ref{App:A}). As we show in App.~\ref{App:B}, the transmission amplitude reads
\begin{equation}
\label{eq:t_QPC}
t_k=\frac{2i\tb t_\textrm{QPC}^{*}\sin k}{|t_\textrm{QPC}|^2-\tb^2 e^{-2ik}}.
\end{equation}
At full transmission, $t_\textrm{QPC}=\tb$, the transmission probability $T=|t_k|^2=1$ is independent of $k$. For the stationary current, we then expect
\begin{equation}
\langle I \rangle =G_QV,
\end{equation}
where $G_Q$ is the conductance quantum ($=1/2\pi$ in our units). This result is shown with a horizontal dashed line in Fig.~\ref{Fig:I_QPC} and agrees well with the quasi-stationary current obtained from our tight-binding calculations. For lower transmission amplitudes, the stationary current reads
\begin{equation}
\label{eq:T_QPC}
\langle I \rangle =G_Q\int_{-V/2}^{V/2}d\eps\,|t_{k(\eps)}|^2,
\end{equation}
which is also confirmed by Fig.~\ref{Fig:I_QPC}. Additionally, we see that our tight-binding calculations of the WTD in Fig.~\ref{Fig:Res_QPC} are in excellent agreement with scattering theory using the transmission amplitudes in Eq.~(\ref{eq:t_QPC}).

\subsection{Single-level quantum dot}
\label{subsec:SLQD}

\begin{figure}
	\includegraphics[width=0.95\columnwidth]{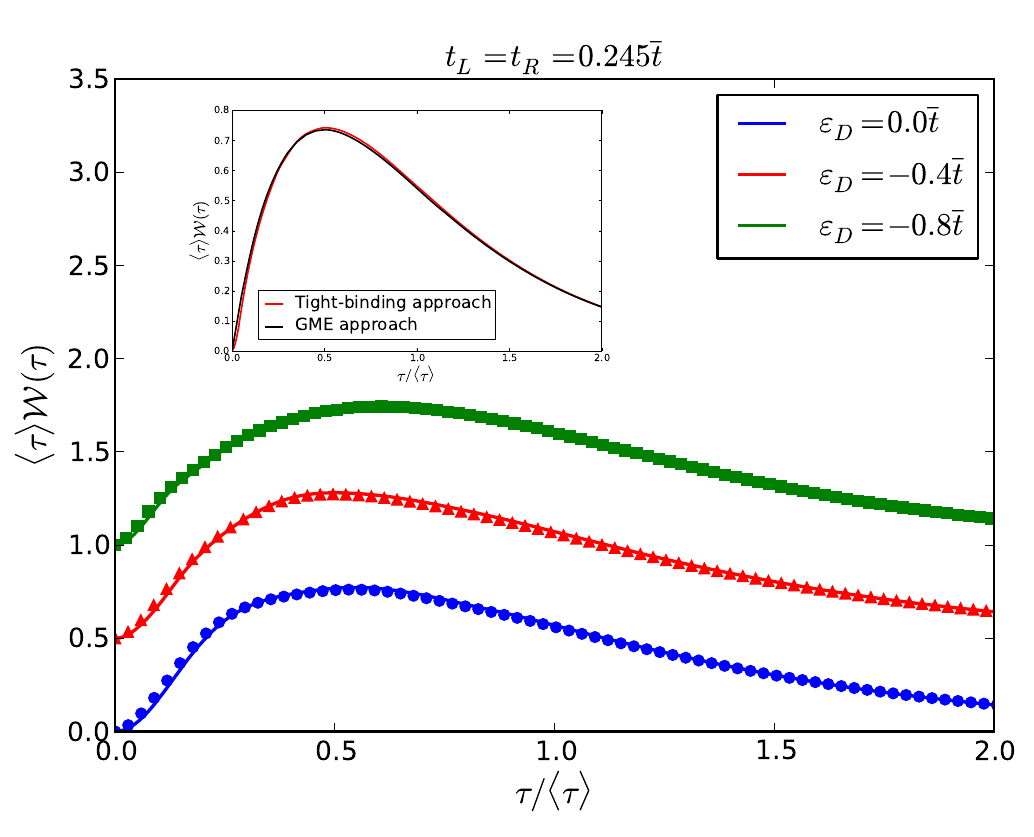}
	\caption{\label{Fig:Res_SD1} (Color online) WTD for a single-level quantum dot. Numerical results (symbols) are shown for varying level positions $\epsilon_D$ and offset vertically for clarity. Calculations based on scattering theory are shown with full lines. The inset shows a comparison between the tight-binding approach (with $\eps_D=0$ and $t_L=t_R=0.15\tb$) and Eq.~(\ref{eq:W_SD}) obtained from a generalized master equation (GME) derived in the high-bias limit.}
\end{figure}

As our next application, we consider a single-level quantum dot. The quantum dot level is denoted as $\ket{D}$ and the corresponding energy is $\epsilon_D$. In this case, the Hamiltonian of the scatterer reads
\begin{equation}
\label{eq:H_sys_SD}
\hatH_\textrm{scat}=\epsilon_D\ket{D}\bra{D},
\end{equation}
whereas the tunneling Hamiltonian takes on the form given by Eq.~(\ref{eq:H_tunnel}).

Figure~\ref{Fig:Res_SD1} shows the WTD for a quantum dot with varying level position $\varepsilon_D$. Again our results agree very well with those obtained from scattering theory. The transmission amplitude is found following the procedure described in App.~\ref{App:B} and reads
\begin{equation}
\label{eq:t_SD}
t_k=\frac{2it_Lt_R\sin k}{(\eps_k-\eps_D)\tb+(t_L^2+t_R^2)e^{ik}}.
\end{equation}
By moving $\varepsilon_D$ away from the center of the energy window, the overall transmission is lowered and the peak of the WTD shifts to larger times.

In a complementary approach, we can calculate the WTD using a GME derived in the high-bias limit following Gurvitz and Prager.\cite{Gurvitz1996} In this approach, the (broadened) energy level is assumed to be positioned well within the energy window $[-V/2,V/2]$. This implies that the rates for tunneling in and out of the level are much slower than the inverse  mean waiting time between incoming particles $\bar \tau=h/V$. We can then effectively set $\bar \tau$ to zero. The tunneling rates from the left lead to the QD ($\alpha=L$) and from the QD to the right lead ($\alpha=R$) are
\begin{equation}
\label{eq:rates}
 \Gamma_{\alpha}\approx \frac{4|t_\alpha|^2}{v_F}
\end{equation}
as shown in App.~\ref{App:C}. With these tunneling rates, the WTD becomes\cite{Brandes2008}
\begin{equation}
\label{eq:W_SD}
\begin{split}
{\cal W}(\tau)&=\frac{\Gamma_R\Gamma_L}{\Gamma_R-\Gamma_L}\of{e^{-\Gamma_L\tau}-e^{-\Gamma_R\tau}}\\
&=\Gamma^2\tau e^{-\Gamma\tau},\,\,\,\, \Gamma_L=\Gamma_R=\Gamma,
\end{split}
\end{equation}
obtained using the GME approach described in App.~\ref{App:D}.

The GME calculations agree very well with our tight-binding results, see inset of Fig.~\ref{Fig:Res_SD1}. There are only small deviations at short waiting times (hardly visible). This is due to the GME approach, which predicts a linear dependence on $\tau$ for $\tau\ll 1/\Gamma$ according to Eq.~(\ref{eq:W_SD}). In contrast, for the tight-binding calculations and from scattering theory, we expect a quadratic dependence on $\tau$ for $\tau\ll\bar{\tau}$.\cite{Albert2012}

\begin{figure}
	\includegraphics[width=0.95\columnwidth]{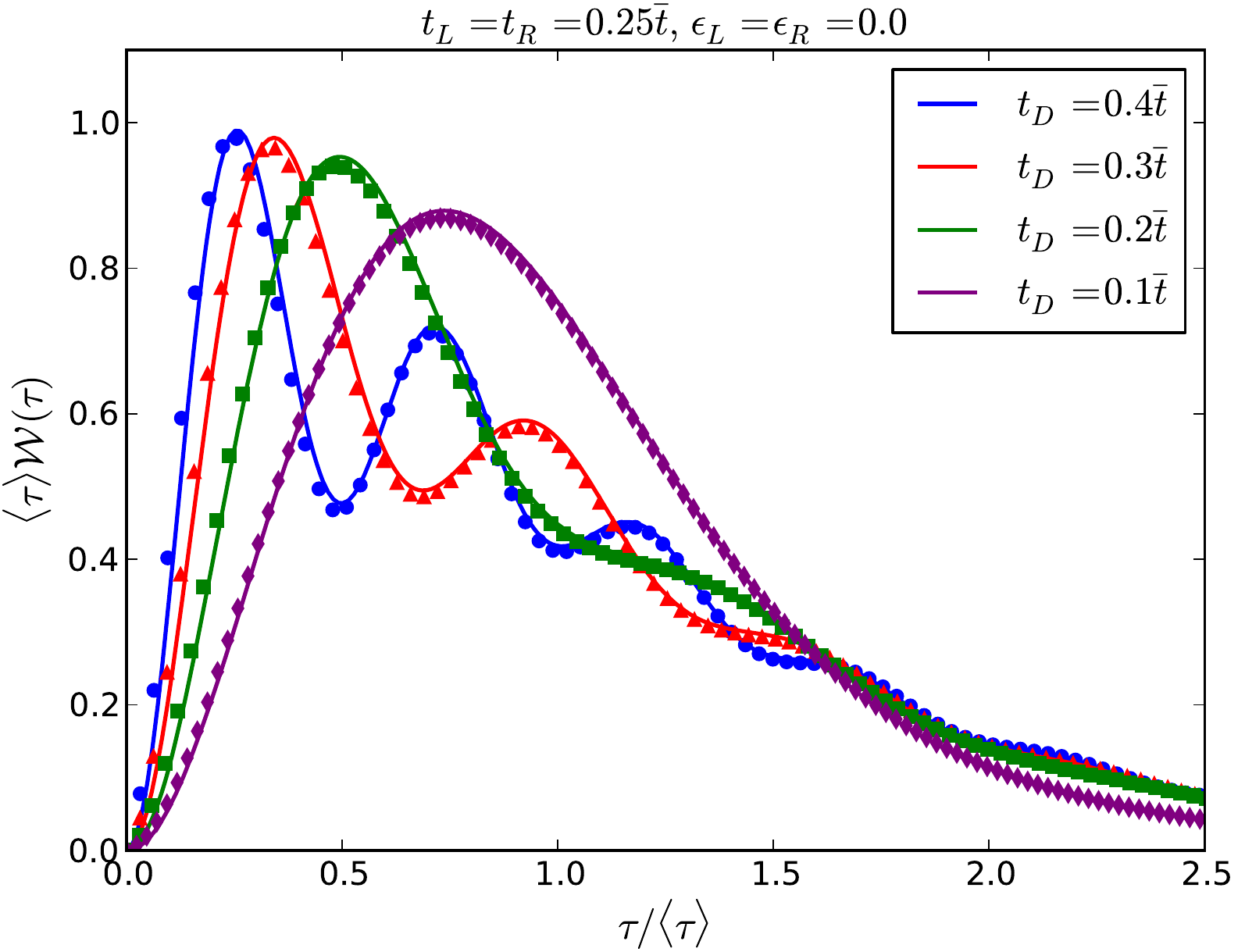}
	\caption{\label{Fig:Res_SDD1}  (Color online) WTD for a serial double quantum dot. Numerical results (symbols) are shown for different interdot coupling strengths $t_D$. Calculations based on scattering theory are shown with full lines.}
\end{figure}

\subsection{Serial double quantum dot}
\label{subsec:SDD}

We now consider a system consisting of two single-level quantum dots in series. The left (right) level $\ket{L}$ ($\ket{R}$) at energy  $\epsilon_L$ ($\epsilon_R$) is coupled to the left (right) lead and the levels are connected by the interdot tunnel coupling $t_D$. The Hamiltonian of the scatterer reads
\begin{equation}
\label{eq:Ham_sDD}
\hatH_\textrm{scat}=\epsilon_L \ket{L} \bra{L} + \epsilon_R \ket{R} \bra{R}-t_D(\ket{L}\bra{R}+\hc)
\end{equation}
and the tunneling Hamiltonian is given by Eq.~(\ref{eq:H_tunnel}),

\begin{figure}
	\includegraphics[width=0.95\columnwidth]{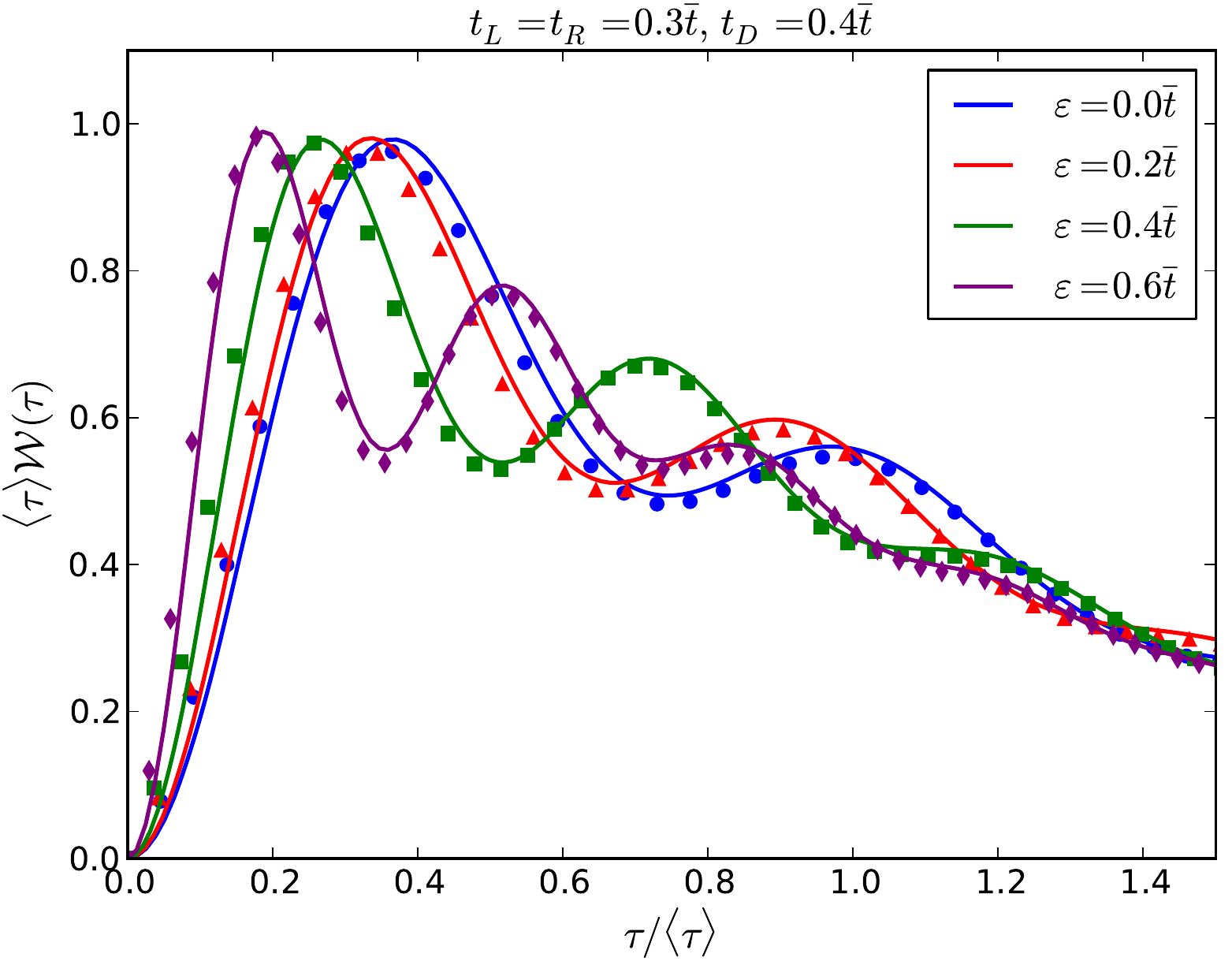}
	\caption{\label{Fig:Res_SDD2} (Color online) WTD for a serial double quantum dot. Numerical results (symbols) are shown for different level separations $\eps=|\eps_L-\eps_R|$. The levels are shifted symmetrically with respect to zero such that $\eps_L=-\eps_R$. Calculations based on scattering theory are shown with full lines.}
\end{figure}

In Fig.~\ref{Fig:Res_SDD1} we show the WTD with different interdot couplings $t_D$ and equal energy levels $\eps_L=\eps_R$. For $t_D>t_L,t_R$, the curves exhibit an oscillatory behavior. As $t_D$ is decreased, the oscillations are damped and the WTD is shifted toward larger times.

To understand the oscillatory behavior we note that the eigenenergies of the Hamiltonian in Eq.~(\ref{eq:Ham_sDD}) are
\begin{equation}
\epsilon_\pm=(\epsilon_L+\epsilon_R)/2 \pm \sqrt{(\epsilon_L-\epsilon_R)^2/4+t_D^2}.
\label{eq:EigenE_DD}
\end{equation}
The difference of the eigenenergies are thus
\begin{equation}
\Delta\epsilon=2\sqrt{\epsilon^2/4+t_D^2},
\label{eq:EigenE_split}
\end{equation}
having defined $\eps=|\eps_L-\eps_R|$. The energy splitting gives rise to coherent oscillations in the WTD with frequency $\omega_{\mathrm{osc}}=\Delta\epsilon$.\cite{Brandes2008,Thomas2013} The oscillations can be understood by noting that when $t_D>t_R$, a particle is likely to oscillate back an forth between the left and right levels before exiting to the right lead. The decay of the WTD at long times is controlled by the tunneling rate to the right lead. To further corroborate this picture, we consider in Fig.~\ref{Fig:Res_SDD2} the WTDs with an increasing detuning of the levels. As expected from Eq.~(\ref{eq:EigenE_split}), the frequency increases as the two levels are dealigned.

For the transmission amplitude, we find in this case
\begin{equation}
\label{eq:t_sDD}
t_k=\frac{2i\tb t_Lt_Dt_R\sin k}{\prod_{\alpha=L,R}[(\eps_k-\eps_\alpha)\tb e^{-ik}+t_\alpha^2]+t_D^2\tb^2e^{-2ik}}.
\end{equation}
Calculations based on scattering theory are in good agreement with our tight-binding calculations as illustrated in Figs.~\ref{Fig:Res_SDD1} and \ref{Fig:Res_SDD2}. We note that the results can also be reproduced (not shown) with high accuracy using the GME approach, see Refs.~\onlinecite{Brandes2008,Thomas2013}.

\subsection{Double quantum dot enclosing a magnetic flux}
\label{subsec:ABDD}

\begin{figure}
\includegraphics[width=0.95\columnwidth]{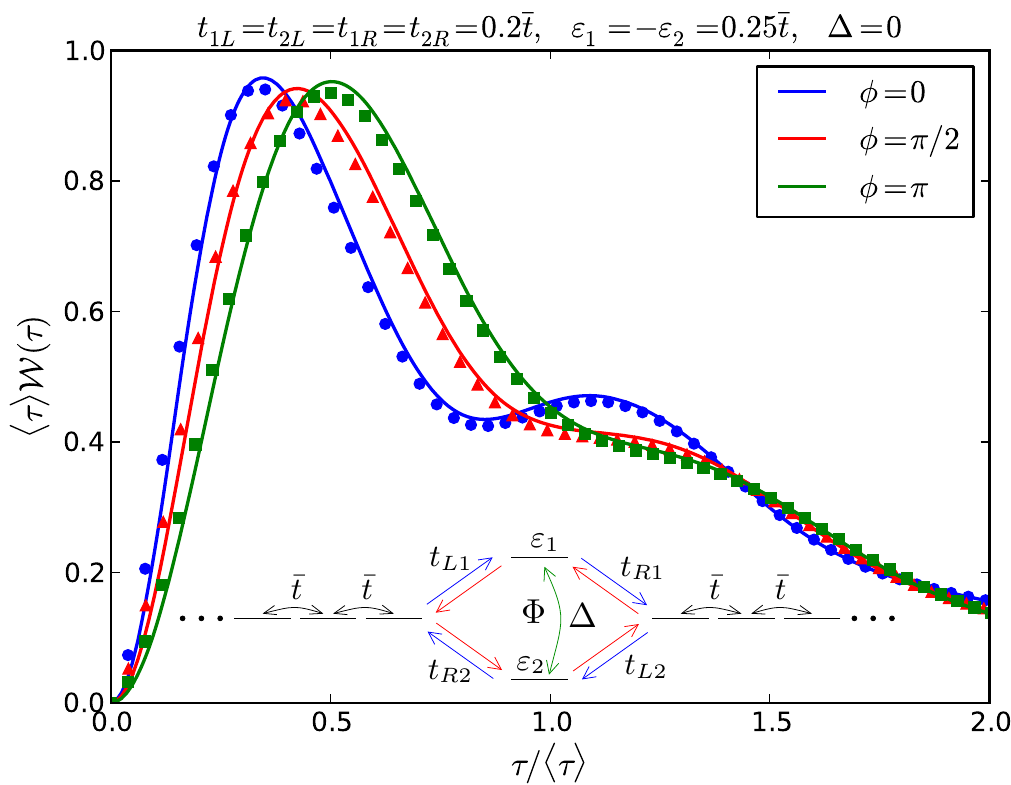}
\caption{\label{Fig:Res_ABDD1}
(Color online) WTD for a parallel double quantum dot enclosing a magnetic flux. The inset shows the setup schematically. A blue (red) arrow implies a phase change of $e^{i\phi/4}$ ($e^{-i\phi/4}$), where $\phi=2\pi(\Phi/\Phi_0)$ is given by the magnetic flux $\Phi$. Direct tunneling between the dots is phase-neutral (green arrows). Tight-binding calculations corresponding to different magnetic fluxes are shown as symbols, results based on scattering theory as solid lines.}
\end{figure}

We now place the two quantum dots in parallel such that each one of them is coupled to both leads. In this setup, a magnetic flux can be enclosed, inducing a variable phase for different paths through the system. The setup is shown schematically in the inset of Fig.~\ref{Fig:Res_ABDD1}. The two levels with energies $\eps_{1(2)}$ are coupled to the left (right) lead by the tunnel couplings $t_{Li}$ ($t_{Ri}$), $i=1,2$. In addition, there is a direct link with tunneling amplitude $\Delta$ between the quantum dots. The magnetic flux $\Phi$ through the central area causes a (charged) particle to acquire a phase factor of $e^{\pm i\phi/4}$ during each hopping event where $\phi=2\pi(\Phi/\Phi_0)$ and $\Phi_0=h/e$ is the magnetic flux quantum.\cite{Welack2009} The plus (minus) sign in the exponential applies if the tunneling event occurs in the clockwise (counterclockwise) direction around $\Phi$.

\begin{widetext}
The Hamiltonian of the double quantum dot now reads
\begin{equation}
\hatH_\textrm{scat}=\eps_1\ket{1}\bra{1}+\eps_2\ket{2}\bra{2}-\Delta(\ket{1}\bra{2}+\ket{2}\bra{1}).
\end{equation}
In addition, the tunneling Hamiltonian is
\begin{eqnarray}
\hatH_\textrm{tun}&=&-t_{L1}\of{e^{i\phi/4}\ket{1}\bra{M_L,L}+\hc}-t_{L2}\of{e^{-i\phi/4}\ket{2}\bra{M_L,L}+\hc}\nonumber\\
&&-t_{R1}\of{e^{-i\phi/4}\ket{1}\bra{1,R}+\hc}-t_{R2}\of{e^{i\phi/4}\ket{2}\bra{1,R}+\hc}.
\end{eqnarray}
For comparison with scattering theory we find for the transmission amplitude
\begin{equation}
\label{eq:t_pDD}
t_k=\frac{2i\tb\sin k \left[t_{L2}t_{R2} (\eps_k-\eps_1)e^{i\phi/2}+t_{L1} t_{R1} (\eps_k-\eps_2)e^{-i\phi/2}-(t_{R1}t_{L2}+t_{L1}t_{R2})\Delta\right]}{|t_{L1}t_{R2}-t_{L2}t_{R1}e^{-i\phi}|^2+\tb e^{-ik}\left[
\sum_{\sigma}(t_{L\sigma}^2+t_{R\sigma}^2)(\eps_k-\eps_{\bar\sigma})-2\Delta\cos(\phi/2)\sum_{\alpha} t_{\alpha1}t_{\alpha2}\right]+\tb^2 e^{-2ik}\left[\prod_{\sigma}(\eps_k-\eps_\sigma)-\Delta^2\right]},
\nonumber
\end{equation}
where $\alpha$ ($=L,R$) is the lead index and, and $\sigma$ ($=1,2$) refers to the quantum dot levels.
\end{widetext}

In Fig.~\ref{Fig:Res_ABDD1} we show WTDs for three different phase shifts $\phi=0$, $\pi/2$, and $\pi$, without direct tunneling between the quantum dots, $\Delta=0$. By varying the phase, we may modify the interference between the two paths leading from the left to the right lead, so that it changes from being constructive ($\phi=0$) to being destructive ($\phi=\pi$). A particle coming from the left lead propagates through both quantum dots and interferes with itself in the right lead. For $\phi=0$, the interference is constructive and particles may perform coherent oscillations as seen in the WTD. At $\phi=\pi$, the interference is maximally destructive and particle transfers through the DQD become increasingly rare. However, because the two paths have different amplitudes (since $\eps_1\neq\eps_2$), the transmission remains non-zero. The reduced transmission decreases the oscillations in the WTD as it approaches an exponential distribution corresponding to a Poisson process.

\begin{figure}
	\includegraphics[width=0.95\columnwidth]{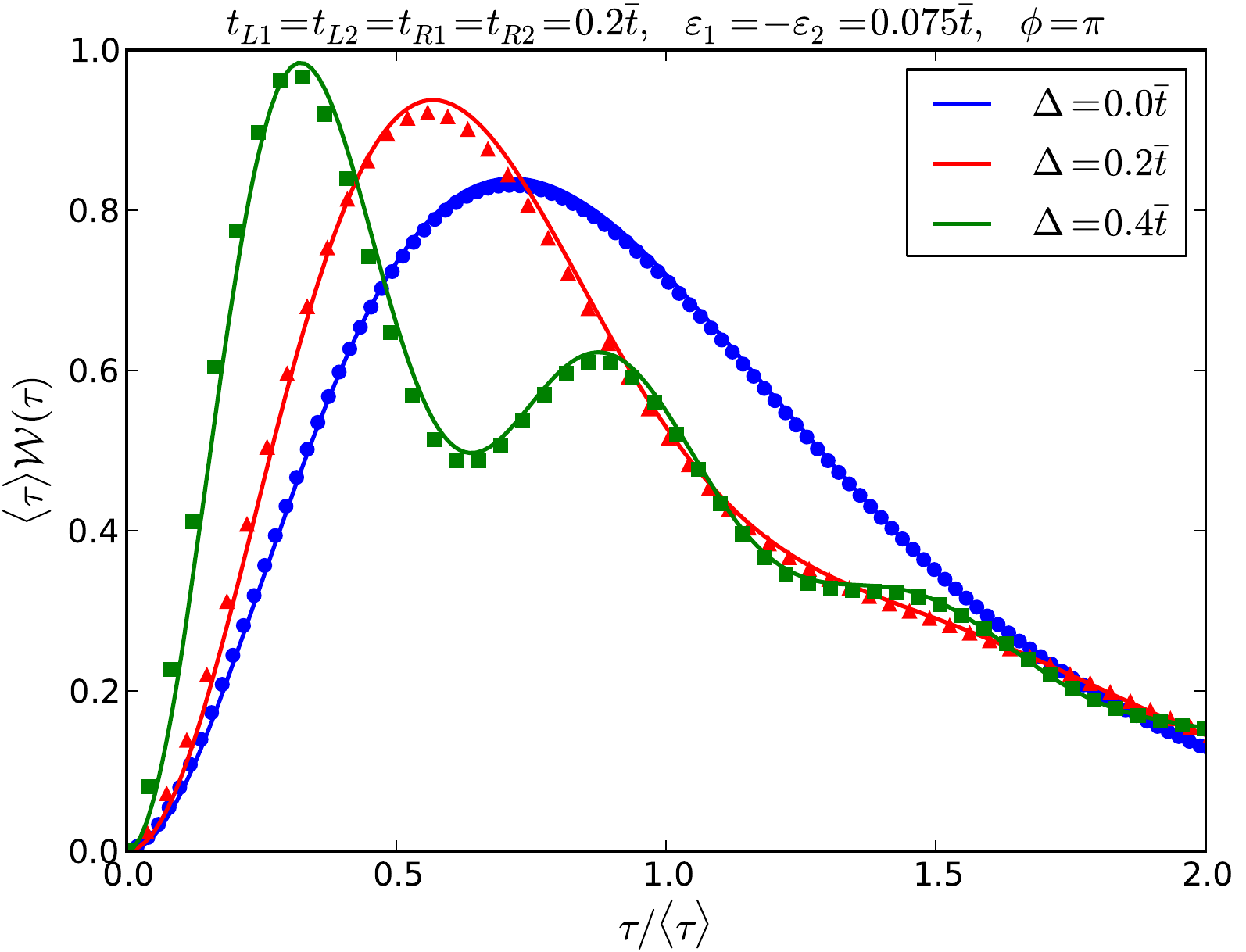}
	\caption{\label{Fig:Res_ABDD2} (Color online) WTD for a parallel double quantum dot enclosing a magnetic flux. The destructive interference between the upper and lower paths is maximal with $\phi=2\pi(\Phi/\Phi_0)=\pi$. By increasing the inter-dot tunneling amplitude $\Delta$, the destructive interference is reduced and the oscillations in the WTD are restored. Tight-binding calculations corresponding to different inter-dot tunneling amplitudes are shown as symbols, results based on scattering theory as solid lines.}
\end{figure}

This picture changes qualitatively with a finite tunneling amplitude between the quantum dots, $\Delta\neq0$, as shown in Fig.~\ref{Fig:Res_ABDD2}, where $\phi=\pi$. Several paths through the systems are now possible so that the interference blockade is lifted and coherent oscillations are restored. In both figures, our tight-binding calculations are in excellent agreement with scattering theory.

\subsection{Bipartite chain}
\label{subsec:Bip}

As a last example we consider transport through a bipartite chain of variable length. This could be a simple model of an extended molecule suspended between two leads.\cite{Su1979, Su1980} The system Hamiltonian
\begin{eqnarray}
\hatH_\textrm{sys}=-\sum_{m=1}^{M_D-1}&&\left\{v\ket{2m-1}\bra{2m}+ w\ket{2m}\bra{2m+1}+\hc\right\}\nonumber\\
&&-\left\{v\ket{2M_D-1}\bra{2M_D}+\hc\right\}
\end{eqnarray}
describes $M_D$ dimers consisting of two sites that are coupled by the tunneling amplitude $v$. Each dimer is in addition coupled to its neighbors with tunneling amplitude $w<v$ and the outermost sites are connected to the leads by the tunneling Hamiltonian in Eq.~(\ref{eq:H_tunnel}).

\begin{figure}
	\includegraphics[width=0.95\columnwidth]{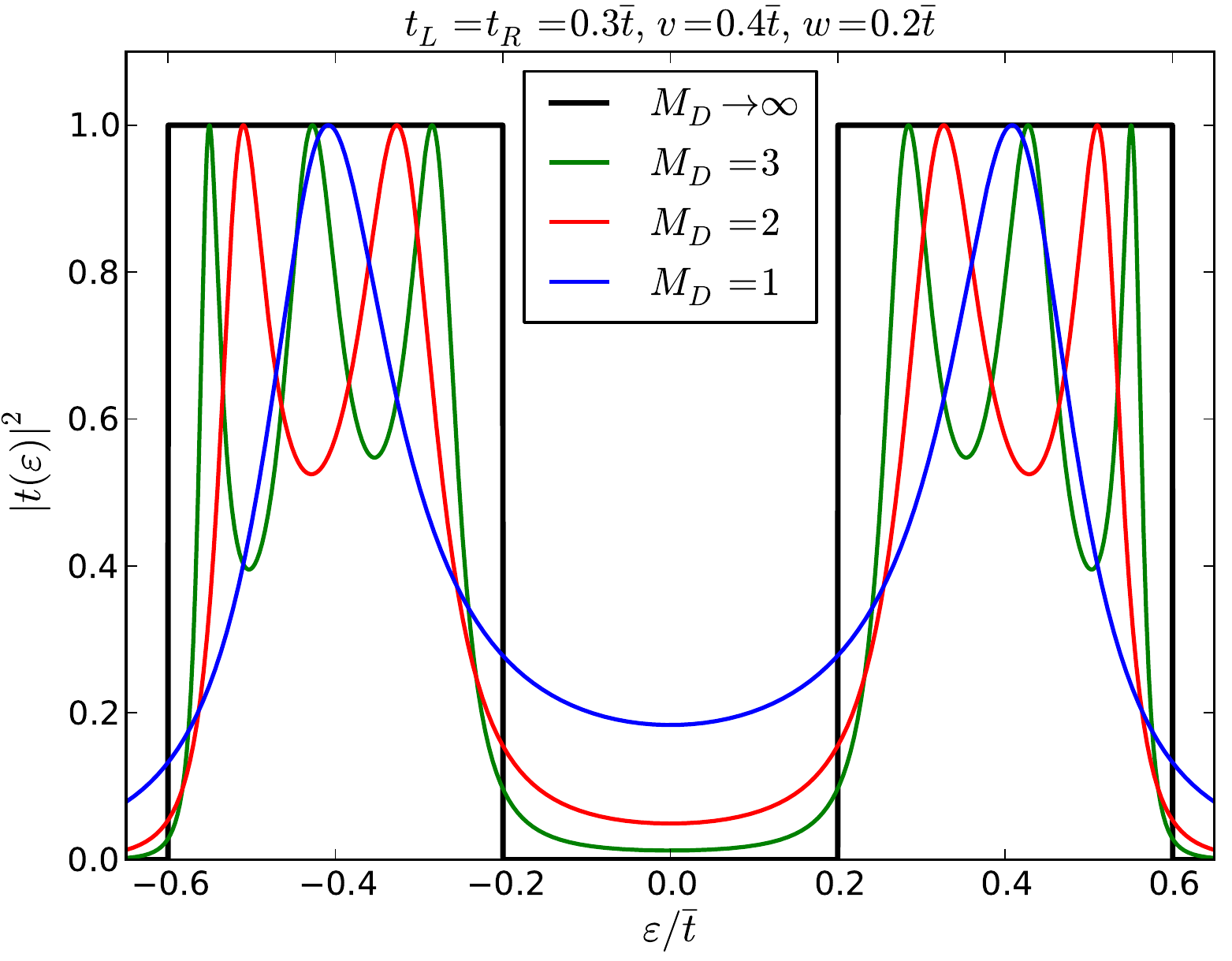}
	\caption{\label{Fig:T_bip} (Color online) Transmission probability for a bipartite chain of length $2M_D$. As the number of dimers $M_D$ increases, the gap around $\eps=0$ becomes clearly defined and the two brands around $\pm v$ become rectangular as indicated with a black line.}
\end{figure}

The transmission amplitudes are obtained numerically for different lengths of the chain using the method described in App.~\ref{App:B}. In Fig.~\ref{Fig:T_bip} we show the energy-dependent transmission obtained for different values of $M_S$. The transmission shows two bands around $\pm v$ with a gap around $\eps=0$ that becomes increasingly pronounced as the length of the chain $2M_D$ is increased. Adding a dimer to the chain increases the number of peaks in the lower and upper bands by one. For $M_D\to\infty$ the peaks become dense within $\pm v-w\leq\eps\leq\pm v+w$ and the transmission peaks become rectangular as shown by the black curve.

\begin{figure}
	\includegraphics[width=0.95\columnwidth]{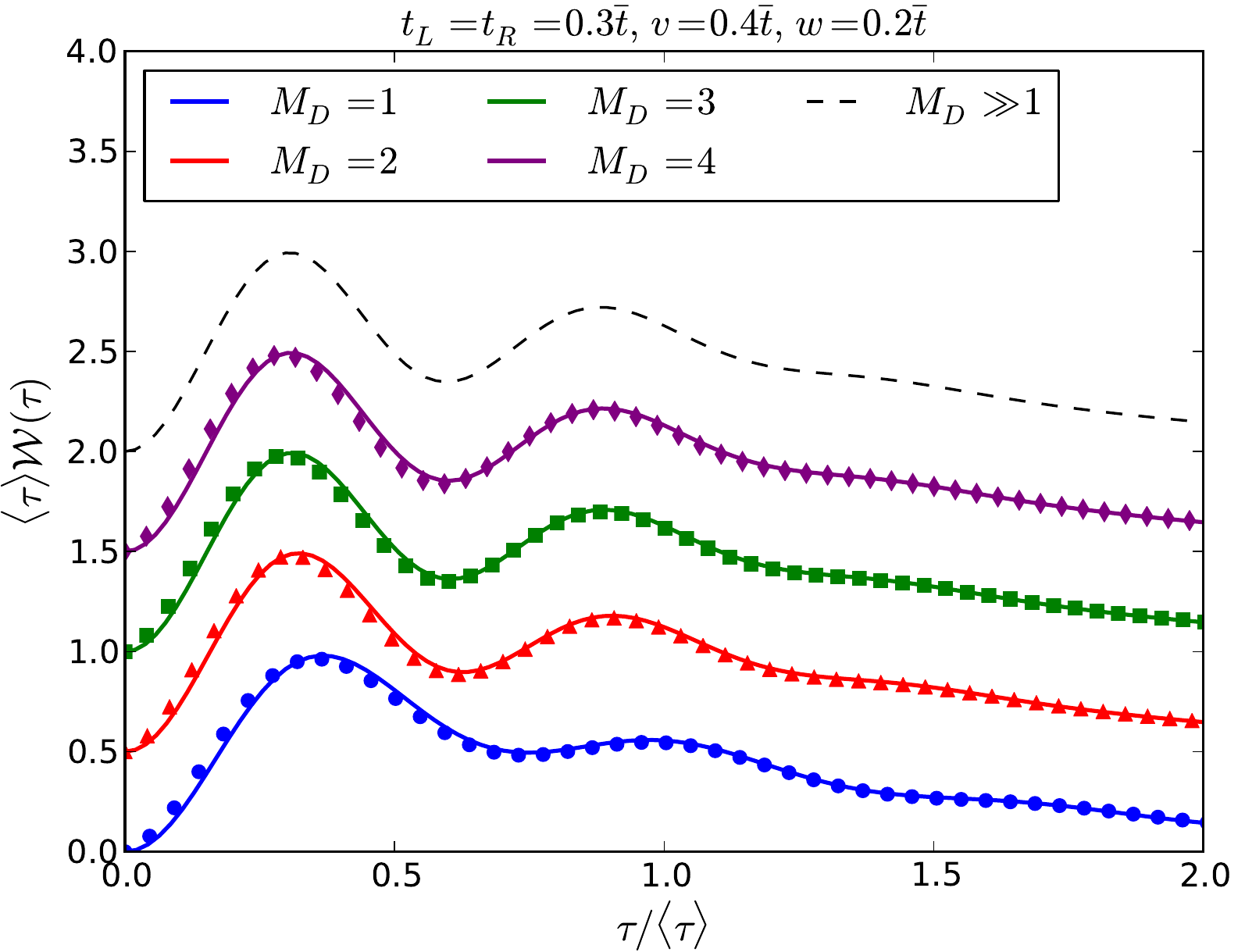}
	\caption{\label{Fig:Res_bip} (Color online) WTD for a bipartite chain of length $2M_D$. Results obtained with the tight-binding method are shown with symbols. Results obtained from scattering theory are indicated with full lines. The curves are offset vertically. As the length of the chain is increased, the results converge towards the universal curve shown with a dashed line. The universal curve is obtained from scattering theory, taking very long chains, where the gap in the spectrum is fully developed.}
\end{figure}

In Fig.~\ref{Fig:Res_bip} we show results for the WTDs for different lengths of the chain. In the case $M_D=1$, we recover the result for a serial double quantum dot with $\eps=0$ and $t_D=v$, cf.~Sec.~\ref{subsec:SDD}. Interestingly, as more dimers are added, the WTDs eventually converge to a universal curve (shown with a dashed line), which is independent of the length $2M_D$.

\section{Conclusions}
\label{sec:Conclusion}

We have presented a method for calculating the waiting time distributions (WTDs) of non-interacting fermions on a finite-size tight-binding chain. As applications of our method, we have calculated the WTDs for a quantum point contact (QPC) and several different quantum dot structures. Our tight-binding approach reproduces the Wigner-Dyson distribution expected for a fully transmitting QPC and it agrees well with predictions based on scattering theory at transmissions below unity. In addition, we can associate oscillations in the WTDs to internal energy scales of quantum dot structures. For quantum dots in series, the oscillations are clearly related to the energy splitting of the hybridized states. For quantum dot structures enclosing a magnetic flux, we find that the WTD carries signatures of the interference between different traversal paths. Finally, for a bipartite chain, the WTDs converge towards a universal curve as the length of the chain is increased. In the high-bias limit, we find good agreement with calculations based on generalized master equations.

The agreement with existing approaches is an important check of our method. In particular, it raises the hope that similar tight-binding calculations may be generalized to include interactions, for example using density matrix renormalization group (DMRG) techniques.\cite{Daley2004,Bohr2006,Carr2011,Chien2012a,Chien2013} It would also be interesting to investigate the WTDs for tight-binding chains with periodic drivings in the spirit of Refs.~\onlinecite{Albert2010,Albert2011,Jonckheere2012,Dasenbrook2013,Albert2014}.

\begin{acknowledgements}
We thank M.~Albert, M.~B\"{u}ttiker, D.~Dasenbrook, G.~Haack, and P.~P.~Hofer for useful discussions and comments. The work was supported by the Swiss National Science Foundation.
\end{acknowledgements}

\appendix

\section{Scattering approach to WTDs}
\label{App:A}

For the sake of completeness, we provide here the essential steps in calculating WTDs within scattering theory following Ref.~\onlinecite{Albert2012}. In this approach, the WTD is calculated in the basis of the scattering states,
\begin{equation}
\varphi_k(x)=
\left\{
\begin{array}{lll}
e^{ikx}+r_ke^{-ikx},& x<0 \\
t_ke^{ikx},& x>x_s>0
\end{array}
\right.
\label{eq:A1}
\end{equation}
where the interval $[0,x_s]$ contains the scatterer with transmission (reflection) amplitudes $t_k$ ($r_k$). The dispersion relation is linearized in the transport window $[\eps_F,\eps_F+eV]$, where $V$ is the applied voltage, such that
\begin{equation}
\epsilon_k=\frac{\hbar^2k^2}{2m}\simeq \eps_F+\hbar v_F k'.
\end{equation}
Here $v_F=\hbar k_F/m$ is the Fermi velocity and we have defined $k'=k-k_F$, which is much smaller than the Fermi momentum, $k'\ll k_F$.

The momentum interval $[k_F,k_F+eV/\hbar v_F]$ is split into $N$ intervals of size $\kappa=eV/N\hbar v_F$. The many-body Slater determinant is constructed from the time-dependent single-particle wave functions
\begin{equation}
\phi_m(x,t)=\frac{e^{-i \eps_F t/\hbar}}{\sqrt{2\pi\kappa}}\int_{\kappa(m-1)}^{\kappa m} dk' e^{-iv_F k't}\varphi_{k_F+k'}(x).
\end{equation}
We moreover define the single-particle operator
\begin{equation}
\hat{\cal Q}_\tau=\int_{x_0}^{x_0+v_F\tau}dx \ket{x}\bra{x},
\end{equation}
where $x_0>x_s$ is located on the right side of the scatterer. The matrix elements of ${\cal Q}_\tau$ are
\begin{equation}
[\mathbf{Q}_\tau]_{m,n}=\bra{\phi_m(\tau)}\hat{\cal Q}_\tau\ket{\phi_n(\tau)},
\end{equation}
which in the limit $N\to\infty$ become
\begin{equation}
[\mathbf{Q}_\tau]_{m,n}=\frac{t_{\kappa m}^* t_{\kappa n}}{2\pi i}\frac{1-e^{iv_F\tau\kappa(n-m)}}{n-m},
\end{equation}
having redefined $t_{k_F+\kappa n}\rightarrow t_{\kappa n}$. Finally, the ITP is \cite{Albert2012}
\begin{equation}
\Pi(\tau)=\det(\mathbf{1}-\mathbf{Q}_\tau)
\end{equation}
from which the WTD follows using Eq.~(\ref{eq:WTD}). We note that only the transmission amplitude $t_{k}$ of the scatterer is required to calculate the WTD.

\section{Transmission amplitudes for tight-binding systems}
\label{App:B}

To obtain the transmission of a given scatterer we consider an incoming plane wave that is transmitted with amplitude $t_k$ and reflected with amplitude~$r_k$, cf.~Eq.~(\ref{eq:A1}).

The Schr\"odinger equation for the eigenstates of the tight-binding Hamiltonian reads
\begin{equation}
\hatH_{\mathrm{tb}}\ket{\phi_k}=\eps_k\ket{\phi_k}.
\end{equation}
We expand the eigenstates on the lattice sites as
\begin{equation}
\ket{\phi_k}=\sum_{\alpha=L,R}\sum_{m=1}^{M_\alpha}c_{\alpha m}^k\ket{m, \alpha}+\sum_{m=1}^{M_s} c_m^k\ket{m},
\end{equation}
where the first sum runs over the sites in the leads and the second sum over the sites of the scatterer.

We evaluate the Schr\"odinger equation on the last site of the left lead and on the first site of the right lead\cite{Feynman1965}
\begin{equation}
\begin{split}
\ew{M_L,L|\hatH|\phi_k}=&\eps_k c_{LM_L}^k=-\tb c_{L(M_L-1)}^k-t_L c_{1}^k,\\
\ew{1,R|\hatH|\phi_k}=&\eps_k c_{R1}^k=-\tb c_{R2}^k-t_R c_{M_s}^k,
\end{split}
\label{eq:siteHtb}
\end{equation}
assuming that the scatterer is coupled to the left (right) lead with hopping amplitude $t_L$ ($t_R$). Similar equations can be formulated for each site of the scatterer, giving us a total of $2+M_s$ equations. Next, we make the ansatz
\begin{equation}
\begin{split}
c_{Lm}^k&=e^{ik(m-M_L)}+r_ke^{-ik(m-M_L)},\\
c_{Rm}^k&=t_ke^{ik(m-1)}
\end{split}
\label{eq:c_Lm}
\end{equation}
for the lead coefficients. Inserting the ansatz into the $2+M_s$ (linear) equations above, we can solve for the amplitudes $t_k$ and $r_k$.

For the QPC considered in Sec.~\ref{subsec:QPC}, Eqs.~(\ref{eq:siteHtb}) and (\ref{eq:c_Lm}) become
\begin{equation}
\begin{split}
\eps_k (1+r_k)&=-\tb (e^{-ik}+r_ke^{ik})-t_\textrm{QPC} t_k,\\
\eps_k t_k&=-\tb t_ke^{ik}-t_\textrm{QPC}^{*} (1+r_k),
\end{split}
\end{equation}
since the leads are directly coupled via the hopping amplitude $t_\textrm{QPC}$. Solving this system of equations, we find
\begin{equation}
t_k=\frac{2i\tb t_\textrm{QPC}^{*}\sin k}{|t_\textrm{QPC}|^2-(\eps_k+\tb e^{ik})^2}
\label{eq:QPC_tk}
\end{equation}
and
\begin{equation}
r_k=\frac{2\tb\eps_k\cos k+\eps_k^2+\tb^2-|t_\textrm{QPC}|^2}{|t_\textrm{QPC}|^2-(\eps_k+\tb e^{ik})^2}.
\end{equation}
It can be verified that $|t_k|^2+|r_k|^2=1$. Moreover, assuming that the dispersion relation $\eps_k=-2\tb\cos k$ still approximately holds, we obtain Eq.~(\ref{eq:t_QPC}) from Eq.~(\ref{eq:QPC_tk}). The transmission amplitudes in Eqs.~(\ref{eq:t_SD}), (\ref{eq:t_sDD}), and (\ref{eq:t_pDD}) as well as in Fig.~\ref{Fig:T_bip} are found in a similar way.

The mean QPC current is now
\begin{equation}
  \langle I\rangle=G_Q\int_{-V/2}^{V/2}d\eps|t_{\eps(k)}|^2,
\end{equation}
where $G_Q$ is the conductance quantum ($=1/2\pi$ in our units). Taking $V=2\tb$ combined with Eq.~(\ref{eq:t_QPC}), we find for the QPC
\begin{equation}
 \langle I\rangle = G_QV\left\{1-\frac{(1-\theta^2)^2}{\theta(1+\theta^2)}\textrm{artanh}\left[\frac\theta{1+\theta^2}\right]\right\},
\end{equation}
where $\theta=t_\textrm{QPC}/\tb$.

\section{Generalized Master equation approach}
\label{App:C}

WTDs can be calculated from GMEs using a method developed by Brandes.\cite{Brandes2008} Here we present a derivation of the WTD for a GME describing uni-directional transport using the language of full counting statistics (FCS).

The scatterer is described by its density matrix $\rhohat_S$ which evolves according to a Markovian GME of the form
\begin{equation}
\frac{d}{dt}\rhohat_S(t)=\mathcal{L}\rhohat_S(t).
\end{equation}
The Liouvillian $\mathcal{L}$ describes the coherent evolution of particles inside the scatterer as well as particle transfers between the scatterer and the leads. To evaluate the FCS we resolve the density matrix with respect to the number of particles $n$ that have been transferred through the scatterer during the time interval $[0,t]$.\cite{Plenio1998,Makhlin2001} From the $n$-resolved density matrix $\rhohat_S^{(n)}(t)$ we obtain the FCS as
\begin{equation}
P(n,t)=\mathrm{Tr}[\rhohat_S^{(n)}(t)].
\end{equation}
Since the transport is assumed to be uni-directional, we have $P(n<0,t)=0$. Additionally, if only one particle at a time can be transferred from the scatterer to the right lead, the GME for the $n$-resolved density matrix reads
\begin{equation}
\frac{d}{dt}\rhohat^{(n)}_S(t)=\mathcal{L}_0\rhohat_S^{(n)}(t)+\mathcal{J}\rhohat_S^{(n-1)}(t),
\label{eq:nresGME}
\end{equation}
having partitioned the Liouvillian as $\mathcal{L}=\mathcal{L}_0+\mathcal{J}$, where the super-operator $\mathcal{J}$ describes individual particle transfers from the scatterer to the lead.

To find the WTD, we use that the idle time probability is simply
\begin{equation}
\Pi(\tau)=P(n=0,\tau)=\mathrm{Tr}[\rhohat_S^{(0)}(\tau)].
\end{equation}
We find $\rhohat_S^{(0)}(\tau)$ by noting that $\rhohat_S^{(-1)}(\tau)=0$, such that Eq.~(\ref{eq:nresGME}) for $n=0$ reduces to
\begin{equation}
\frac{d}{dt}\rhohat^{(0)}_S(t)=\mathcal{L}_0\rhohat_S^{(0)}(t).
\end{equation}
The formal solution for $\rhohat_S^{(0)}(t)$ is then
\begin{equation}
\rhohat_S^{(0)}(t)=e^{{\cal L}_0t}\rhohat_S^\textrm{stat},
\end{equation}
assuming that the system (prepared in an arbitrary state in the distant past) has reached the stationary state $\rhohat_S^\textrm{stat}$ at $t=0$. The stationary state is obtained as the normalized solution to
\begin{equation}
{\cal L}\rhohat_S^\textrm{stat}=0.
\label{eq:stat_state}
\end{equation}
The idle time probability is now
\begin{equation}
\Pi(\tau)=\mathrm{Tr}[e^{{\cal L}_0\tau}\rhohat_S^\textrm{stat}].
\end{equation}
From the idle time probability we first obtain the mean waiting time using Eq.~(\ref{eq:meanWT})
\begin{equation}
\ew{\tau}=-\frac{1}{\dot \Pi(\tau=0)}=-\frac{1}{\mathrm{Tr}[{\cal L}_0\rho_S^\textrm{stat}]}=\frac{1}{\mathrm{Tr}[{\cal J}\rhohat_S^\textrm{stat}]}.
\end{equation}
Here we have used that ${\cal L}_0\rho_S^\textrm{stat}=({\cal L}-\mathcal{J})\rho_S^\textrm{stat}=-\mathcal{J}\rho_S^\textrm{stat}$, since ${\cal L}\rho_S^\textrm{stat}=0$ according to Eq.~(\ref{eq:stat_state}). The mean waiting time is simply the inverse average particle current.

Finally, the WTD follows from Eq.~(\ref{eq:WTD}) as
\begin{equation}
{\cal W}(\tau)=\ew{\tau}\partial_\tau^2 \Pi(\tau)=\frac{\mathrm{Tr}\left[{\cal J}e^{{\cal L}_0\tau}{\cal J}\rhohat_S^\textrm{stat}\right]}{\mathrm{Tr}[{\cal J}\rhohat_S^\textrm{stat}]}.
\label{eq:WTD_Brandes}
\end{equation}
in agreement with the result by Brandes.\cite{Brandes2008} Here we have used that $\mathrm{Tr}[{\cal L}_0\bullet]=\mathrm{Tr}[({\cal L}-\mathcal{J})\bullet]=-\mathrm{Tr}[\mathcal{J}\bullet]$, since $\mathrm{Tr}[\mathcal{L}\bullet]=0$ due to probability conservation.

As an illustration, we consider the single-level quantum dot from section \ref{subsec:SLQD}. Here, it suffices to consider the diagonal elements of the density matrix, which we denote by $\rhohat_0$ (empty level) and $\rhohat_1$ (full level). Combining the elements into the vector $\rhohat_S=(\rhohat_0,\rhohat_1)^T$, the two parts of the Liouvillian ${\cal L}={\cal L}_{0}+{\cal J}$ take the matrix forms
\begin{equation}
{\cal L}_{0}=\left(\begin{array}{cc} -\Gamma_L & 0 \\ \Gamma_L & -\Gamma_R \end{array}\right),\hspace{0.5cm}
{\cal J}=\left(\begin{array}{cc} 0 & \Gamma_R \\ 0 & 0 \end{array}\right).
\end{equation}
The rates $\Gamma_L$ and $\Gamma_R$ can be expressed in terms of the tight-binding parameters as described in App.~\ref{App:D}. From Eq.~(\ref{eq:WTD_Brandes}) then follows the WTD in Eq.~(\ref{eq:W_SD}).

\section{Derivation of transition rates}
\label{App:D}

We evaluate the rates entering the GME from our tight-binding model. The Hamiltonians of the leads in Eq.~(\ref{eq:H_alpha}) can be diagonalized by the transformation
\begin{equation}
\label{eq:FTrans}
\ket{m,\alpha}=\sqrt{\frac{2}{M_\alpha+1}}\sum_{j=1}^{M_\alpha}\sin(k_\alpha^j m)\ket{k_\alpha^j},\hspace{0.2cm}
k_\alpha^j=\frac{j\pi}{M_\alpha+1},
\end{equation}
with $j=1\ldots,M_\alpha$, leading to the eigenenergies
\begin{equation}
\eps_{k_\alpha^j}=-2\tb\cos(k_\alpha^j).
\label{eq:eigenenergies}
\end{equation}
We take a generic tunneling Hamiltonian
\begin{equation}
\label{eq:V_gen}\hatH_T = -\sum_{\alpha=L,R}\sum_{\mu=1}^{N_s}\of{t_{\alpha\mu} \ket{\mu}\bra{m_\alpha,\alpha}+t_{\alpha\mu}^*\ket{m_\alpha,\alpha}\bra{\mu}}
\end{equation}
with $m_L=M_L$ and $m_R=1$, connecting the outermost sites of the leads to the $N_s$ sites of the scatterer. Applying the transformation in Eq.~(\ref{eq:FTrans}) to the tunneling Hamiltonian then yields
\begin{equation}
\hatH_T = -\sum_{\alpha=L,R} \sum_{j=1}^{M_\alpha} \sum_{\mu=1}^{N_s}\of{t_{\alpha\mu}^j\ket{\mu}\bra{k_\alpha^j}
+(t_{\alpha\mu}^j)^*\ket{k_\alpha^j}\bra{\mu}}
\end{equation}
with the hopping amplitudes
\begin{equation}
\label{eq:t_k_alpha}
t_{\alpha\mu}^j=\sqrt{\frac{2}{M_\alpha+1}}\sin(k_\alpha^j)t_{\alpha\mu}.
\end{equation}

With the level of the scatterer well inside the energy window $[-V/2,V/2]$, the particle transport is unidirectional and the transition rates between the leads and the scatterer are
\begin{equation}
\Gamma_{\alpha\mu}(\eps)=2\pi\sum_{j=1}^{M_\alpha} |t_{\alpha\mu}^j|^2\delta(\eps-\eps_{\alpha}^j).
\end{equation}
From Eq.~(\ref{eq:t_k_alpha}) we then get
\begin{equation}
\begin{split}
\Gamma_{\alpha\mu}(\eps)=&\frac{4\pi |t_{\alpha\mu}|^2}{M_\alpha+1} \sum_{j=1}^{M_\alpha}\sin^2(k_\alpha^j)\delta(\eps-\eps_\alpha^{j})\\
\approx&\frac{4\pi |t_{\alpha\mu}|^2}{M_\alpha+1}\frac{M_\alpha}{\pi} \int_0^\pi dk_\alpha\sin^2(k_\alpha)\delta(\eps-\eps_{k_\alpha}).
\end{split}
\end{equation}
Moreover, using Eq.~(\ref{eq:eigenenergies}) and taking $M_\alpha\gg 1$, we find
\begin{equation}
\label{eq:TR_alpha_mu}
\Gamma_{\alpha\mu}(\eps)=\frac{2 |t_{\alpha\mu}|^2}{\tb}\sqrt{1-\of{\frac\eps{2\tb}}^2}.	
\end{equation}
Around the center of the band ($\eps\simeq 0$), this gives
\begin{equation}
\label{eq:Gamma_const}
\Gamma_{\alpha\mu}(\eps\simeq 0)\approx\frac{2 |t_{\alpha\mu}|^2}{\tb}=\frac{4 |t_{\alpha\mu}|^2}{v_F}.
\end{equation}


\begin{thebibliography}{100}%
\makeatletter
\providecommand \@ifxundefined [1]{%
 \@ifx{#1\undefined}
}%
\providecommand \@ifnum [1]{%
 \ifnum #1\expandafter \@firstoftwo
 \else \expandafter \@secondoftwo
 \fi
}%
\providecommand \@ifx [1]{%
 \ifx #1\expandafter \@firstoftwo
 \else \expandafter \@secondoftwo
 \fi
}%
\providecommand \natexlab [1]{#1}%
\providecommand \enquote  [1]{``#1''}%
\providecommand \bibnamefont  [1]{#1}%
\providecommand \bibfnamefont [1]{#1}%
\providecommand \citenamefont [1]{#1}%
\providecommand \href@noop [0]{\@secondoftwo}%
\providecommand \href [0]{\begingroup \@sanitize@url \@href}%
\providecommand \@href[1]{\@@startlink{#1}\@@href}%
\providecommand \@@href[1]{\endgroup#1\@@endlink}%
\providecommand \@sanitize@url [0]{\catcode `\\12\catcode `\$12\catcode
  `\&12\catcode `\#12\catcode `\^12\catcode `\_12\catcode `\%12\relax}%
\providecommand \@@startlink[1]{}%
\providecommand \@@endlink[0]{}%
\providecommand \url  [0]{\begingroup\@sanitize@url \@url }%
\providecommand \@url [1]{\endgroup\@href {#1}{\urlprefix }}%
\providecommand \urlprefix  [0]{URL }%
\providecommand \Eprint [0]{\href }%
\@ifxundefined \urlstyle {%
  \providecommand \doi  [0]{\begingroup \@sanitize@url \@doi}%
  \providecommand \@doi [1]{\endgroup \@@startlink {\doibase
  #1}doi:\discretionary {}{}{}#1\@@endlink }%
}{%
  \providecommand \doi  [0]{doi:\discretionary{}{}{}\begingroup
  \urlstyle{rm}\Url }%
}%
\providecommand \doibase [0]{http://dx.doi.org/}%
\providecommand \Doi [0]{\begingroup \@sanitize@url \@Doi }%
\providecommand \@Doi  [1]{\endgroup\@@startlink{\doibase#1}\@@Doi}%
\providecommand \@@Doi [1]{#1\@@endlink}%
\providecommand \selectlanguage [0]{\@gobble}%
\providecommand \bibinfo  [0]{\@secondoftwo}%
\providecommand \bibfield  [0]{\@secondoftwo}%
\providecommand \translation [1]{[#1]}%
\providecommand \BibitemOpen [0]{}%
\providecommand \bibitemStop [0]{}%
\providecommand \bibitemNoStop [0]{.\EOS\space}%
\providecommand \EOS [0]{\spacefactor3000\relax}%
\providecommand \BibitemShut  [1]{\csname bibitem#1\endcsname}%
\bibitem [{\citenamefont {Gabelli}\ \emph {et~al.}(2006)\citenamefont
  {Gabelli}, \citenamefont {F\`{e}ve}, \citenamefont {Berroir}, \citenamefont
  {Pla\c{c}ais}, \citenamefont {Cavanna}, \citenamefont {Etienne},
  \citenamefont {Jin},\ and\ \citenamefont {Glattli}}]{Gabelli2006}%
  \BibitemOpen
  \bibfield  {author} {\bibinfo {author} {\bibfnamefont {J.}~\bibnamefont
  {Gabelli}}, \bibinfo {author} {\bibfnamefont {G.}~\bibnamefont {F\`{e}ve}},
  \bibinfo {author} {\bibfnamefont {J.-M.}\ \bibnamefont {Berroir}}, \bibinfo
  {author} {\bibfnamefont {B.}~\bibnamefont {Pla\c{c}ais}}, \bibinfo {author}
  {\bibfnamefont {A.}~\bibnamefont {Cavanna}}, \bibinfo {author} {\bibfnamefont
  {B.}~\bibnamefont {Etienne}}, \bibinfo {author} {\bibfnamefont
  {Y.}~\bibnamefont {Jin}}, \ and\ \bibinfo {author} {\bibfnamefont {D.~C.}\
  \bibnamefont {Glattli}},\ }\Doi {10.1126/science.1126940} {\bibfield
  {journal} {\bibinfo  {journal} {Science} }\textbf {\bibinfo {volume}
  {313}},\ \bibinfo {pages} {499} (\bibinfo {year} {2006})}\BibitemShut
  {NoStop}%
\bibitem [{\citenamefont {F\`{e}ve}\ \emph {et~al.}(2007)\citenamefont
  {F\`{e}ve}, \citenamefont {Mah\'e}, \citenamefont {Berroir}, \citenamefont
  {Kontos}, \citenamefont {Pla\c{c}ais}, \citenamefont {Glattli}, \citenamefont
  {Cavanna}, \citenamefont {Etienne},\ and\ \citenamefont {Jin}}]{Feve2007}%
  \BibitemOpen
  \bibfield  {author} {\bibinfo {author} {\bibfnamefont {G.}~\bibnamefont
  {F\`{e}ve}}, \bibinfo {author} {\bibfnamefont {A.}~\bibnamefont {Mah\'e}},
  \bibinfo {author} {\bibfnamefont {J.-M.}\ \bibnamefont {Berroir}}, \bibinfo
  {author} {\bibfnamefont {T.}~\bibnamefont {Kontos}}, \bibinfo {author}
  {\bibfnamefont {B.}~\bibnamefont {Pla\c{c}ais}}, \bibinfo {author}
  {\bibfnamefont {D.~C.}\ \bibnamefont {Glattli}}, \bibinfo {author}
  {\bibfnamefont {A.}~\bibnamefont {Cavanna}}, \bibinfo {author} {\bibfnamefont
  {B.}~\bibnamefont {Etienne}}, \ and\ \bibinfo {author} {\bibfnamefont
  {Y.}~\bibnamefont {Jin}},\ }\Doi {10.1126/science.1141243} {\bibfield
  {journal} {\bibinfo  {journal} {Science} }\textbf {\bibinfo {volume}
  {316}},\ \bibinfo {pages} {1169} (\bibinfo {year} {2007})}\BibitemShut
  {NoStop}%
\bibitem [{\citenamefont {Bocquillon}\ \emph {et~al.}(2012)\citenamefont
  {Bocquillon}, \citenamefont {Parmentier}, \citenamefont {Grenier},
  \citenamefont {Berroir}, \citenamefont {Degiovanni}, \citenamefont {Glattli},
  \citenamefont {Pla\ifmmode~\mbox{\c{c}}\else \c{c}\fi{}ais}, \citenamefont
  {Cavanna}, \citenamefont {Jin},\ and\ \citenamefont
  {F\`eve}}]{Bocquillon2012}%
  \BibitemOpen
  \bibfield  {author} {\bibinfo {author} {\bibfnamefont {E.}~\bibnamefont
  {Bocquillon}}, \bibinfo {author} {\bibfnamefont {F.~D.}\ \bibnamefont
  {Parmentier}}, \bibinfo {author} {\bibfnamefont {C.}~\bibnamefont {Grenier}},
  \bibinfo {author} {\bibfnamefont {J.-M.}\ \bibnamefont {Berroir}}, \bibinfo
  {author} {\bibfnamefont {P.}~\bibnamefont {Degiovanni}}, \bibinfo {author}
  {\bibfnamefont {D.~C.}\ \bibnamefont {Glattli}}, \bibinfo {author}
  {\bibfnamefont {B.}~\bibnamefont {Pla\ifmmode~\mbox{\c{c}}\else
  \c{c}\fi{}ais}}, \bibinfo {author} {\bibfnamefont {A.}~\bibnamefont
  {Cavanna}}, \bibinfo {author} {\bibfnamefont {Y.}~\bibnamefont {Jin}}, \ and\
  \bibinfo {author} {\bibfnamefont {G.}~\bibnamefont {F\`eve}},\ }\Doi
  {10.1103/PhysRevLett.108.196803} {\bibfield  {journal} {\bibinfo  {journal}
  {Phys. Rev. Lett.} }\textbf {\bibinfo {volume} {108}},\ \bibinfo {pages}
  {196803} (\bibinfo {year} {2012})}\BibitemShut {NoStop}%
\bibitem [{\citenamefont {Bocquillon}\ \emph {et~al.}(2014)\citenamefont
  {Bocquillon}, \citenamefont {Freulon}, \citenamefont {Parmentier},
  \citenamefont {Berroir}, \citenamefont {Pla\c{c}ais}, \citenamefont {Wahl},
  \citenamefont {Rech}, \citenamefont {Jonckheere}, \citenamefont {Martin},
  \citenamefont {Grenier}, \citenamefont {Ferraro}, \citenamefont
  {Degiovanni},\ and\ \citenamefont {F\`{e}ve}}]{Bocquillon2013}%
  \BibitemOpen
  \bibfield  {author} {\bibinfo {author} {\bibfnamefont {E.}~\bibnamefont
  {Bocquillon}}, \bibinfo {author} {\bibfnamefont {V.}~\bibnamefont {Freulon}},
  \bibinfo {author} {\bibfnamefont {F.~D.}\ \bibnamefont {Parmentier}},
  \bibinfo {author} {\bibfnamefont {J.-M.}\ \bibnamefont {Berroir}}, \bibinfo
  {author} {\bibfnamefont {B.}~\bibnamefont {Pla\c{c}ais}}, \bibinfo {author}
  {\bibfnamefont {C.}~\bibnamefont {Wahl}}, \bibinfo {author} {\bibfnamefont
  {J.}~\bibnamefont {Rech}}, \bibinfo {author} {\bibfnamefont {T.}~\bibnamefont
  {Jonckheere}}, \bibinfo {author} {\bibfnamefont {T.}~\bibnamefont {Martin}},
  \bibinfo {author} {\bibfnamefont {C.}~\bibnamefont {Grenier}}, \bibinfo
  {author} {\bibfnamefont {D.}~\bibnamefont {Ferraro}}, \bibinfo {author}
  {\bibfnamefont {P.}~\bibnamefont {Degiovanni}}, \ and\ \bibinfo {author}
  {\bibfnamefont {G.}~\bibnamefont {F\`{e}ve}},\ }\Doi {10.1002/andp.201300181}
  {\bibfield  {journal} {\bibinfo  {journal} {Ann. Phys. (Berlin)} }\textbf
  {\bibinfo {volume} {526}},\ \bibinfo {pages} {1} (\bibinfo {year}
  {2014})}\BibitemShut {NoStop}%
\bibitem [{\citenamefont {Fletcher}\ \emph {et~al.}(2013)\citenamefont
  {Fletcher}, \citenamefont {See}, \citenamefont {Howe}, \citenamefont
  {Pepper}, \citenamefont {Giblin}, \citenamefont {Griffiths}, \citenamefont
  {Jones}, \citenamefont {Farrer}, \citenamefont {Ritchie}, \citenamefont
  {Janssen},\ and\ \citenamefont {Kataoka}}]{Fletcher2013}%
  \BibitemOpen
  \bibfield  {author} {\bibinfo {author} {\bibfnamefont {J.~D.}\ \bibnamefont
  {Fletcher}}, \bibinfo {author} {\bibfnamefont {P.}~\bibnamefont {See}},
  \bibinfo {author} {\bibfnamefont {H.}~\bibnamefont {Howe}}, \bibinfo {author}
  {\bibfnamefont {M.}~\bibnamefont {Pepper}}, \bibinfo {author} {\bibfnamefont
  {S.~P.}\ \bibnamefont {Giblin}}, \bibinfo {author} {\bibfnamefont {J.~P.}\
  \bibnamefont {Griffiths}}, \bibinfo {author} {\bibfnamefont {G.~A.~C.}\
  \bibnamefont {Jones}}, \bibinfo {author} {\bibfnamefont {I.}~\bibnamefont
  {Farrer}}, \bibinfo {author} {\bibfnamefont {D.~A.}\ \bibnamefont {Ritchie}},
  \bibinfo {author} {\bibfnamefont {T.~J. B.~M.}\ \bibnamefont {Janssen}}, \
  and\ \bibinfo {author} {\bibfnamefont {M.}~\bibnamefont {Kataoka}},\ }\Doi
  {10.1103/PhysRevLett.111.216807} {\bibfield  {journal} {\bibinfo  {journal}
  {Phys. Rev. Lett.} }\textbf {\bibinfo {volume} {111}},\ \bibinfo {pages}
  {216807} (\bibinfo {year} {2013})}\BibitemShut {NoStop}%
\bibitem [{\citenamefont {Dubois}\ \emph {et~al.}(2013)\citenamefont {Dubois},
  \citenamefont {Jullien}, \citenamefont {Portier}, \citenamefont {Roche},
  \citenamefont {Cavanna}, \citenamefont {Jin}, \citenamefont {Wegscheider},
  \citenamefont {Roulleau},\ and\ \citenamefont {Glattli}}]{Dubois2013}%
  \BibitemOpen
  \bibfield  {author} {\bibinfo {author} {\bibfnamefont {J.}~\bibnamefont
  {Dubois}}, \bibinfo {author} {\bibfnamefont {T.}~\bibnamefont {Jullien}},
  \bibinfo {author} {\bibfnamefont {F.}~\bibnamefont {Portier}}, \bibinfo
  {author} {\bibfnamefont {P.}~\bibnamefont {Roche}}, \bibinfo {author}
  {\bibfnamefont {A.}~\bibnamefont {Cavanna}}, \bibinfo {author} {\bibfnamefont
  {Y.}~\bibnamefont {Jin}}, \bibinfo {author} {\bibfnamefont {W.}~\bibnamefont
  {Wegscheider}}, \bibinfo {author} {\bibfnamefont {P.}~\bibnamefont
  {Roulleau}}, \ and\ \bibinfo {author} {\bibfnamefont {D.~C.}\ \bibnamefont
  {Glattli}},\ }\Doi {10.1038/nature12713} {\bibfield  {journal} {\bibinfo
  {journal} {Nature} }\textbf {\bibinfo {volume} {502}},\ \bibinfo {pages}
  {659} (\bibinfo {year} {2013})}\BibitemShut {NoStop}%
\bibitem [{\citenamefont {Gaury}\ \emph {et~al.}(2014)\citenamefont {Gaury},
  \citenamefont {Weston}, \citenamefont {Santin}, \citenamefont {Houzet},
  \citenamefont {Groth},\ and\ \citenamefont {Waintal}}]{Gaury2014}%
  \BibitemOpen
  \bibfield  {author} {\bibinfo {author} {\bibfnamefont {B.}~\bibnamefont
  {Gaury}}, \bibinfo {author} {\bibfnamefont {J.}~\bibnamefont {Weston}},
  \bibinfo {author} {\bibfnamefont {M.}~\bibnamefont {Santin}}, \bibinfo
  {author} {\bibfnamefont {M.}~\bibnamefont {Houzet}}, \bibinfo {author}
  {\bibfnamefont {C.}~\bibnamefont {Groth}}, \ and\ \bibinfo {author}
  {\bibfnamefont {X.}~\bibnamefont {Waintal}},\ }\Doi
  {http://dx.doi.org/10.1016/j.physrep.2013.09.001} {\bibfield  {journal}
  {\bibinfo  {journal} {Phys. Rep.} }\textbf {\bibinfo {volume} {534}},\
  \bibinfo {pages} {1} (\bibinfo {year} {2014})}\BibitemShut {NoStop}%
\bibitem [{\citenamefont {Blanter}\ and\ \citenamefont
  {B\"{u}ttiker}(2000)}]{Blanter2000}%
  \BibitemOpen
  \bibfield  {author} {\bibinfo {author} {\bibfnamefont {Ya.~M.}\ \bibnamefont
  {Blanter}}\ and\ \bibinfo {author} {\bibfnamefont {M.}~\bibnamefont
  {B\"{u}ttiker}},\ }\href@noop {} {\bibfield  {journal} {\bibinfo  {journal}
  {Phys. Rep.} }\textbf {\bibinfo {volume} {336}},\ \bibinfo {pages} {1}
  (\bibinfo {year} {2000})}\BibitemShut {NoStop}%
\bibitem [{\citenamefont {Reulet}\ \emph {et~al.}(2003)\citenamefont {Reulet},
  \citenamefont {Senzier},\ and\ \citenamefont {Prober}}]{Reulet2003}%
  \BibitemOpen
  \bibfield  {author} {\bibinfo {author} {\bibfnamefont {B.}~\bibnamefont
  {Reulet}}, \bibinfo {author} {\bibfnamefont {J.}~\bibnamefont {Senzier}}, \
  and\ \bibinfo {author} {\bibfnamefont {D.~E.}\ \bibnamefont {Prober}},\ }\Doi
  {10.1103/PhysRevLett.91.196601} {\bibfield  {journal} {\bibinfo  {journal}
  {Phys. Rev. Lett.} }\textbf {\bibinfo {volume} {91}},\ \bibinfo {pages}
  {196601} (\bibinfo {year} {2003})}\BibitemShut {NoStop}%
\bibitem [{\citenamefont {Bomze}\ \emph {et~al.}(2005)\citenamefont {Bomze},
  \citenamefont {Gershon}, \citenamefont {Shovkun}, \citenamefont {Levitov},\
  and\ \citenamefont {Reznikov}}]{Bomze2005}%
  \BibitemOpen
  \bibfield  {author} {\bibinfo {author} {\bibfnamefont {Yu.}~\bibnamefont
  {Bomze}}, \bibinfo {author} {\bibfnamefont {G.}~\bibnamefont {Gershon}},
  \bibinfo {author} {\bibfnamefont {D.}~\bibnamefont {Shovkun}}, \bibinfo
  {author} {\bibfnamefont {L.~S.}\ \bibnamefont {Levitov}}, \ and\ \bibinfo
  {author} {\bibfnamefont {M.}~\bibnamefont {Reznikov}},\ }\Doi
  {10.1103/PhysRevLett.95.176601} {\bibfield  {journal} {\bibinfo  {journal}
  {Phys. Rev. Lett.} }\textbf {\bibinfo {volume} {95}},\ \bibinfo {pages}
  {176601} (\bibinfo {year} {2005})}\BibitemShut {NoStop}%
\bibitem [{\citenamefont {Gershon}\ \emph {et~al.}(2008)\citenamefont
  {Gershon}, \citenamefont {Bomze}, \citenamefont {Sukhorukov},\ and\
  \citenamefont {Reznikov}}]{Gershon2007}%
  \BibitemOpen
  \bibfield  {author} {\bibinfo {author} {\bibfnamefont {G.}~\bibnamefont
  {Gershon}}, \bibinfo {author} {\bibfnamefont {Yu.}~\bibnamefont {Bomze}},
  \bibinfo {author} {\bibfnamefont {E.~V.}\ \bibnamefont {Sukhorukov}}, \ and\
  \bibinfo {author} {\bibfnamefont {M.}~\bibnamefont {Reznikov}},\ }\Doi
  {10.1103/PhysRevLett.101.016803} {\bibfield  {journal} {\bibinfo  {journal}
  {Phys. Rev. Lett.} }\textbf {\bibinfo {volume} {101}},\ \bibinfo {pages}
  {016803} (\bibinfo {year} {2008})}\BibitemShut {NoStop}%
\bibitem [{\citenamefont {Timofeev}\ \emph {et~al.}(2007)\citenamefont
  {Timofeev}, \citenamefont {Meschke}, \citenamefont {Peltonen}, \citenamefont
  {Heikkila},\ and\ \citenamefont {Pekola}}]{Timofeev2007}%
  \BibitemOpen
  \bibfield  {author} {\bibinfo {author} {\bibfnamefont {A.~V.}\ \bibnamefont
  {Timofeev}}, \bibinfo {author} {\bibfnamefont {M.}~\bibnamefont {Meschke}},
  \bibinfo {author} {\bibfnamefont {J.~T.}\ \bibnamefont {Peltonen}}, \bibinfo
  {author} {\bibfnamefont {T.~T.}\ \bibnamefont {Heikkila}}, \ and\ \bibinfo
  {author} {\bibfnamefont {J.~P.}\ \bibnamefont {Pekola}},\ }\Doi
  {10.1103/PhysRevLett.98.207001}
  {\bibfield  {journal} {\bibinfo  {journal} {Phys Rev. Lett.} }\textbf
  {\bibinfo {volume} {98}},\ \bibinfo {pages} {207001} (\bibinfo {year}
  {2007})}\BibitemShut {NoStop}%
\bibitem [{\citenamefont {Levitov}\ and\ \citenamefont
  {Lesovik}(1993)}]{Levitov1993}%
  \BibitemOpen
  \bibfield  {author} {\bibinfo {author} {\bibfnamefont {L.~S.}\ \bibnamefont
  {Levitov}}\ and\ \bibinfo {author} {\bibfnamefont {G.~B.}\ \bibnamefont
  {Lesovik}},\ }\href@noop {} {\bibfield  {journal} {\bibinfo  {journal} {JETP
  Lett.} }\textbf {\bibinfo {volume} {58}},\ \bibinfo {pages} {230 }
  (\bibinfo {year} {1993})}\BibitemShut {NoStop}%
\bibitem [{\citenamefont {Levitov}\ \emph {et~al.}(1996)\citenamefont
  {Levitov}, \citenamefont {Lee},\ and\ \citenamefont {Lesovik}}]{Levitov1996}%
  \BibitemOpen
  \bibfield  {author} {\bibinfo {author} {\bibfnamefont {L.~S.}\ \bibnamefont
  {Levitov}}, \bibinfo {author} {\bibfnamefont {H.}~\bibnamefont {Lee}}, \ and\
  \bibinfo {author} {\bibfnamefont {G.~B.}\ \bibnamefont {Lesovik}},\ }\Doi
  {10.1063/1.531672} {\bibfield  {journal} {\bibinfo  {journal} {J. Math.
  Phys.} }\textbf {\bibinfo {volume} {37}},\ \bibinfo {pages} {4845}
  (\bibinfo {year} {1996})}\BibitemShut {NoStop}%
\bibitem [{\citenamefont {Nazarov}(2003)}]{Nazarov2003}%
  \BibitemOpen
  \bibinfo {editor} {\bibfnamefont {Yu.~V.}\ \bibnamefont {Nazarov}},\ ed.,\
  \href@noop {} {\emph {\bibinfo {title} {Quantum Noise in Mesoscopic
  Physics}}}\ (\bibinfo  {publisher} {Kluwer, Dordrecht},\ \bibinfo {year}
  {2003})\BibitemShut {NoStop}%
\bibitem[{\citenamefont{Gustavsson et~al.}(2006)\citenamefont{Gustavsson,
  Leturcq, Simovic, Schleser, Ihn, Studerus, Ensslin, Driscoll, and
  Gossard}}]{Gustavsson2006}
\bibinfo{author}{\bibfnamefont{S.}~\bibnamefont{Gustavsson}},
  \bibinfo{author}{\bibfnamefont{R.}~\bibnamefont{Leturcq}},
  \bibinfo{author}{\bibfnamefont{B.}~\bibnamefont{Simovic}},
  \bibinfo{author}{\bibfnamefont{R.}~\bibnamefont{Schleser}},
  \bibinfo{author}{\bibfnamefont{T.}~\bibnamefont{Ihn}},
  \bibinfo{author}{\bibfnamefont{P.}~\bibnamefont{Studerus}},
  \bibinfo{author}{\bibfnamefont{K.}~\bibnamefont{Ensslin}},
  \bibinfo{author}{\bibfnamefont{D.~C.} \bibnamefont{Driscoll}},
  \bibnamefont{and} \bibinfo{author}{\bibfnamefont{A.~C.}
  \bibnamefont{Gossard}}, \bibinfo{journal}{Phys. Rev. Lett.}
  \textbf{\bibinfo{volume}{96}}, \bibinfo{pages}{076605}
  (\bibinfo{year}{2006}).
\bibitem[{\citenamefont{Sukhorukov et~al.}(2007)\citenamefont{Sukhorukov,
  Jordan, Gustavsson, Leturcq, Ihn, and Ensslin}}]{Sukhorukov2007}
\bibinfo{author}{\bibfnamefont{E.~V.} \bibnamefont{Sukhorukov}},
  \bibinfo{author}{\bibfnamefont{A.~N.} \bibnamefont{Jordan}},
  \bibinfo{author}{\bibfnamefont{S.}~\bibnamefont{Gustavsson}},
  \bibinfo{author}{\bibfnamefont{R.}~\bibnamefont{Leturcq}},
  \bibinfo{author}{\bibfnamefont{T.}~\bibnamefont{Ihn}}, \bibnamefont{and}
  \bibinfo{author}{\bibfnamefont{K.}~\bibnamefont{Ensslin}},
  \bibinfo{journal}{Nature Physics} \textbf{\bibinfo{volume}{3}},
  \bibinfo{pages}{243} (\bibinfo{year}{2007}).

\bibitem[{\citenamefont{Gustavsson et~al.}(2009)\citenamefont{Gustavsson,
  Leturcq, Studer, Shorubalko, Ihn, Ensslin, Driscoll, and
  Gossard}}]{Gustavsson2009}
\bibinfo{author}{\bibfnamefont{S.}~\bibnamefont{Gustavsson}},
  \bibinfo{author}{\bibfnamefont{R.}~\bibnamefont{Leturcq}},
  \bibinfo{author}{\bibfnamefont{M.}~\bibnamefont{Studer}},
  \bibinfo{author}{\bibfnamefont{I.}~\bibnamefont{Shorubalko}},
  \bibinfo{author}{\bibfnamefont{T.}~\bibnamefont{Ihn}},
  \bibinfo{author}{\bibfnamefont{K.}~\bibnamefont{Ensslin}},
  \bibinfo{author}{\bibfnamefont{D.~C.} \bibnamefont{Driscoll}},
  \bibnamefont{and} \bibinfo{author}{\bibfnamefont{A.~C.}
  \bibnamefont{Gossard}}, \bibinfo{journal}{Surf. Sci. Rep.}
  \textbf{\bibinfo{volume}{64}}, \bibinfo{pages}{191} (\bibinfo{year}{2009}).

\bibitem[{\citenamefont{Flindt et~al.}(2009)\citenamefont{Flindt, Fricke,
  Hohls, Novotn\'{y}, Neto\v{c}n\'{y}, Brandes, and Haug}}]{Flindt2009}
\bibinfo{author}{\bibfnamefont{C.}~\bibnamefont{Flindt}},
  \bibinfo{author}{\bibfnamefont{C.}~\bibnamefont{Fricke}},
  \bibinfo{author}{\bibfnamefont{F.}~\bibnamefont{Hohls}},
  \bibinfo{author}{\bibfnamefont{T.}~\bibnamefont{Novotn\'{y}}},
  \bibinfo{author}{\bibfnamefont{K.}~\bibnamefont{Neto\v{c}n\'{y}}},
  \bibinfo{author}{\bibfnamefont{T.}~\bibnamefont{Brandes}}, \bibnamefont{and}
  \bibinfo{author}{\bibfnamefont{R.~J.} \bibnamefont{Haug}},
  \bibinfo{journal}{Proc. Natl. Acad. Sci. USA} \textbf{\bibinfo{volume}{106}},
  \bibinfo{pages}{10116} (\bibinfo{year}{2009}).

\bibitem[{\citenamefont{Fricke et~al.}(2010{\natexlab{a}})\citenamefont{Fricke,
  Hohls, Flindt, and Haug}}]{Fricke2010a}
\bibinfo{author}{\bibfnamefont{C.}~\bibnamefont{Fricke}},
  \bibinfo{author}{\bibfnamefont{F.}~\bibnamefont{Hohls}},
  \bibinfo{author}{\bibfnamefont{C.}~\bibnamefont{Flindt}}, \bibnamefont{and}
  \bibinfo{author}{\bibfnamefont{R.~J.} \bibnamefont{Haug}},
  \bibinfo{journal}{Physica E} \textbf{\bibinfo{volume}{42}},
  \bibinfo{pages}{848} (\bibinfo{year}{2010}{\natexlab{a}}).
\bibitem[{\citenamefont{Fricke et~al.}(2010{\natexlab{b}})\citenamefont{Fricke,
  Hohls, Sethubalasubramanian, Fricke, and Haug}}]{Fricke2010b}
\bibinfo{author}{\bibfnamefont{C.}~\bibnamefont{Fricke}},
  \bibinfo{author}{\bibfnamefont{F.}~\bibnamefont{Hohls}},
  \bibinfo{author}{\bibfnamefont{N.}~\bibnamefont{Sethubalasubramanian}},
  \bibinfo{author}{\bibfnamefont{L.}~\bibnamefont{Fricke}}, \bibnamefont{and}
  \bibinfo{author}{\bibfnamefont{R.~J.} \bibnamefont{Haug}},
  \bibinfo{journal}{Appl. Phys. Lett.} \textbf{\bibinfo{volume}{96}},
  \bibinfo{pages}{202103} (\bibinfo{year}{2010}{\natexlab{b}}).
\bibitem [{\citenamefont {Ubbelohde}\ \emph
  {et~al.}(2012){\natexlab{b}}\citenamefont {Ubbelohde}, \citenamefont
  {Fricke}, \citenamefont {Flindt}, \citenamefont {Hohls},\ and\ \citenamefont
  {Haug}}]{Ubbelohde2012}%
  \BibitemOpen
  \bibfield  {author} {\bibinfo {author} {\bibfnamefont {N.}~\bibnamefont
  {Ubbelohde}},\ {\bibfnamefont {C.}~\bibnamefont
  {Fricke}},\ {\bibfnamefont {C.}~\bibnamefont
  {Flindt}},\ {\bibfnamefont {F.}~\bibnamefont
  {Hohls}},\  and\ {\bibfnamefont {R.~J.}~\bibnamefont
  {Haug}},\ }\Doi {10.1038/ncomms1620}
  {\bibfield  {journal} {\bibinfo  {journal} {Nat. Commun.} }\textbf
  {\bibinfo {volume} {3}},\ \bibinfo {pages} {612} (\bibinfo {year}
  {2012}{\natexlab{b}})}\BibitemShut {NoStop}%
\bibitem [{\citenamefont {Albert}\ \emph {et~al.}(2014)\citenamefont {Maisi},
  \citenamefont {Kambly}, \citenamefont {Flindt},\ and\ \citenamefont
  {Pekola}}]{Maisi2014}%
  \BibitemOpen
  \bibfield  {author} {\bibinfo {author} {\bibfnamefont {V. F.}~\bibnamefont
  {Maisi}}, \bibinfo {author} {\bibfnamefont {D.}~\bibnamefont {Kambly}},
  \bibinfo {author} {\bibfnamefont {C.}~\bibnamefont {Flindt}}, \ and\ \bibinfo
  {author} {\bibfnamefont {J.~P.}~\bibnamefont {Pekola}},\ }\Doi
  {10.1103/PhysRevLett.112.036801} {\bibfield  {journal} {\bibinfo  {journal}
  {Phys. Rev. Lett.} }\textbf {\bibinfo {volume} {112}},\ \bibinfo {pages}
  {036801} (\bibinfo {year} {2014})}\BibitemShut {NoStop}%
\bibitem [{\citenamefont {Hassler}\ \emph {et~al.}(2008)\citenamefont
  {Hassler}, \citenamefont {Suslov}, \citenamefont {Graf}, \citenamefont
  {Lebedev}, \citenamefont {Lesovik},\ and\ \citenamefont
  {Blatter}}]{Hassler2008}%
  \BibitemOpen
  \bibfield  {author} {\bibinfo {author} {\bibfnamefont {F.}~\bibnamefont
  {Hassler}}, \bibinfo {author} {\bibfnamefont {M.~V.}\ \bibnamefont {Suslov}},
  \bibinfo {author} {\bibfnamefont {G.~M.}\ \bibnamefont {Graf}}, \bibinfo
  {author} {\bibfnamefont {M.~V.}\ \bibnamefont {Lebedev}}, \bibinfo {author}
  {\bibfnamefont {G.~B.}\ \bibnamefont {Lesovik}}, \ and\ \bibinfo {author}
  {\bibfnamefont {G.}~\bibnamefont {Blatter}},\ }\Doi
  {10.1103/PhysRevB.78.165330} {\bibfield  {journal} {\bibinfo  {journal}
  {Phys. Rev. B} }\textbf {\bibinfo {volume} {78}},\ \bibinfo {pages}
  {165330} (\bibinfo {year} {2008})}\BibitemShut {NoStop}%
\bibitem [{\citenamefont {Brandes}(2008)}]{Brandes2008}%
  \BibitemOpen
  \bibfield  {author} {\bibinfo {author} {\bibfnamefont {T.}~\bibnamefont
  {Brandes}},\ }\Doi {10.1002/andp.200810306} {\bibfield  {journal} {\bibinfo
  {journal} {Ann. Phys. (Berlin)} }\textbf {\bibinfo {volume} {17}},\
  \bibinfo {pages} {477} (\bibinfo {year} {2008})}\BibitemShut {NoStop}%
\bibitem [{\citenamefont {Welack}\ \emph {et~al.}(2009)\citenamefont {Welack},
  \citenamefont {Mukamel},\ and\ \citenamefont {Yan}}]{Welack2009}%
  \BibitemOpen
  \bibfield  {author} {\bibinfo {author} {\bibfnamefont {S.}~\bibnamefont
  {Welack}}, \bibinfo {author} {\bibfnamefont {S.}~\bibnamefont {Mukamel}}, \
  and\ \bibinfo {author} {\bibfnamefont {Y.}~\bibnamefont {Yan}},\ }\Doi
  {10.1209/0295-5075/85/57008} {\bibfield  {journal} {\bibinfo  {journal}
  {Europhys. Lett.} }\textbf {\bibinfo {volume} {85}},\ \bibinfo {pages}
  {57008} (\bibinfo {year} {2009})}\BibitemShut {NoStop}%
\bibitem [{\citenamefont {Albert}\ \emph {et~al.}(2011)\citenamefont {Albert},
  \citenamefont {Flindt},\ and\ \citenamefont {B\"uttiker}}]{Albert2011}%
  \BibitemOpen
  \bibfield  {author} {\bibinfo {author} {\bibfnamefont {M.}~\bibnamefont
  {Albert}}, \bibinfo {author} {\bibfnamefont {C.}~\bibnamefont {Flindt}}, \
  and\ \bibinfo {author} {\bibfnamefont {M.}~\bibnamefont {B\"uttiker}},\ }\Doi
  {10.1103/PhysRevLett.107.086805} {\bibfield  {journal} {\bibinfo  {journal}
  {Phys. Rev. Lett.} }\textbf {\bibinfo {volume} {107}},\ \bibinfo {pages}
  {086805} (\bibinfo {year} {2011})}\BibitemShut {NoStop}%
\bibitem [{\citenamefont {Albert}\ \emph {et~al.}(2012)\citenamefont {Albert},
  \citenamefont {Haack}, \citenamefont {Flindt},\ and\ \citenamefont
  {B\"uttiker}}]{Albert2012}%
  \BibitemOpen
  \bibfield  {author} {\bibinfo {author} {\bibfnamefont {M.}~\bibnamefont
  {Albert}}, \bibinfo {author} {\bibfnamefont {G.}~\bibnamefont {Haack}},
  \bibinfo {author} {\bibfnamefont {C.}~\bibnamefont {Flindt}}, \ and\ \bibinfo
  {author} {\bibfnamefont {M.}~\bibnamefont {B\"uttiker}},\ }\Doi
  {10.1103/PhysRevLett.108.186806} {\bibfield  {journal} {\bibinfo  {journal}
  {Phys. Rev. Lett.} }\textbf {\bibinfo {volume} {108}},\ \bibinfo {pages}
  {186806} (\bibinfo {year} {2012})}\BibitemShut {NoStop}%
\bibitem [{\citenamefont {Thomas}\ and\ \citenamefont
  {Flindt}(2013)}]{Thomas2013}%
  \BibitemOpen
  \bibfield  {author} {\bibinfo {author} {\bibfnamefont {K.~H.}\ \bibnamefont
  {Thomas}}\ and\ \bibinfo {author} {\bibfnamefont {C.}~\bibnamefont
  {Flindt}},\ }\Doi {10.1103/PhysRevB.87.121405} {\bibfield  {journal}
  {\bibinfo  {journal} {Phys. Rev. B} }\textbf {\bibinfo {volume} {87}},\
  \bibinfo {pages} {121405(R)} (\bibinfo {year} {2013})}\BibitemShut {NoStop}%
\bibitem [{\citenamefont {Rajabi}\ \emph {et~al.}(2013)\citenamefont {Rajabi},
  \citenamefont {P\"oltl},\ and\ \citenamefont {Governale}}]{Rajabi2013}%
  \BibitemOpen
  \bibfield  {author} {\bibinfo {author} {\bibfnamefont {L.}~\bibnamefont
  {Rajabi}}, \bibinfo {author} {\bibfnamefont {C.}~\bibnamefont {P\"oltl}}, \
  and\ \bibinfo {author} {\bibfnamefont {M.}~\bibnamefont {Governale}},\ }\Doi
  {10.1103/PhysRevLett.111.067002} {\bibfield  {journal} {\bibinfo  {journal}
  {Phys. Rev. Lett.} }\textbf {\bibinfo {volume} {111}},\ \bibinfo {pages}
  {067002} (\bibinfo {year} {2013})}\BibitemShut {NoStop}%
\bibitem [{\citenamefont {Dasenbrook}\ \emph {et~al.}(2013)\citenamefont
  {Dasenbrook}, \citenamefont {Flindt},\ and\ \citenamefont
  {B\"{u}ttiker}}]{Dasenbrook2013}%
  \BibitemOpen
  \bibfield  {author} {\bibinfo {author} {\bibfnamefont {D.}~\bibnamefont
  {Dasenbrook}}, \bibinfo {author} {\bibfnamefont {C.}~\bibnamefont {Flindt}},
  \ and\ \bibinfo {author} {\bibfnamefont {M.}~\bibnamefont {B\"{u}ttiker}},\
  }\Doi
  {10.1103/PhysRevLett.112.146801}
  {\bibfield  {journal} {\bibinfo  {journal}
  {Phys. Rev. Lett.} }\textbf {\bibinfo {volume} {112}},\ \bibinfo {pages}
  {146801} (\bibinfo {year} {2014})}\BibitemShut {NoStop}%
\bibitem [{\citenamefont {Albert}\ and\ \citenamefont
  {Devillard}()}]{Albert2014}%
  \BibitemOpen
  \bibfield  {author} {\bibinfo {author} {\bibfnamefont {M.}~\bibnamefont
  {Albert}}\ and\ \bibinfo {author} {\bibfnamefont {P.}~\bibnamefont
  {Devillard}},\ }\href@noop {} {\bibinfo  {journal}
  {arXiv:1401.5723}}\BibitemShut {NoStop}%
\bibitem [{\citenamefont {Vyas}\ and\ \citenamefont {Singh}(1988)}]{Vyas1988}%
  \BibitemOpen
\bibfield  {journal} {  }\bibfield  {author} {\bibinfo {author} {\bibfnamefont
  {R.}~\bibnamefont {Vyas}}\ and\ \bibinfo {author} {\bibfnamefont
  {S.}~\bibnamefont {Singh}},\ }\Doi {10.1103/PhysRevA.38.2423} {\bibfield
  {journal} {\bibinfo  {journal} {Phys. Rev. A} }\textbf {\bibinfo {volume}
  {38}},\ \bibinfo {pages} {2423} (\bibinfo {year} {1988})}\BibitemShut
  {NoStop}%
\bibitem [{\citenamefont {Carmichael}\ \emph {et~al.}(1989)\citenamefont
  {Carmichael}, \citenamefont {Singh}, \citenamefont {Vyas},\ and\
  \citenamefont {Rice}}]{Carmichael1989}%
  \BibitemOpen
  \bibfield  {author} {\bibinfo {author} {\bibfnamefont {H.~J.}\ \bibnamefont
  {Carmichael}}, \bibinfo {author} {\bibfnamefont {S.}~\bibnamefont {Singh}},
  \bibinfo {author} {\bibfnamefont {R.}~\bibnamefont {Vyas}}, \ and\ \bibinfo
  {author} {\bibfnamefont {P.~R.}\ \bibnamefont {Rice}},\ }\Doi
  {10.1103/PhysRevA.39.1200} {\bibfield  {journal} {\bibinfo  {journal} {Phys.
  Rev. A} }\textbf {\bibinfo {volume} {39}},\ \bibinfo {pages} {1200}
  (\bibinfo {year} {1989})}\BibitemShut {NoStop}%
\bibitem [{\citenamefont {Sch\"onhammer}(2007)}]{Schonhammer2007}%
  \BibitemOpen
  \bibfield  {author} {\bibinfo {author} {\bibfnamefont {K.}~\bibnamefont
  {Sch\"onhammer}},\ }\Doi {10.1103/PhysRevB.75.205329} {\bibfield  {journal}
  {\bibinfo  {journal} {Phys. Rev. B} }\textbf {\bibinfo {volume} {75}},\
  \bibinfo {pages} {205329} (\bibinfo {year} {2007})}\BibitemShut {NoStop}%
\bibitem [{\citenamefont {Inhester}\ and\ \citenamefont
  {Sch\"onhammer}(2009)}]{Inhester2009}%
  \BibitemOpen
  \bibfield  {author} {\bibinfo {author} {\bibfnamefont {L.}~\bibnamefont
  {Inhester}}\ and\ \bibinfo {author} {\bibfnamefont {K.}~\bibnamefont
  {Sch\"onhammer}},\ }\Doi {10.1088/0953-8984/21/47/474209} {\bibfield
  {journal} {\bibinfo  {journal} {J. Phys.: Condens. Matter} }\textbf
  {\bibinfo {volume} {21}},\ \bibinfo {pages} {474209} (\bibinfo {year}
  {2009})}\BibitemShut {NoStop}%
\bibitem [{\citenamefont {Sch\"onhammer}(2009)}]{Schonhammer2009}%
  \BibitemOpen
  \bibfield  {author} {\bibinfo {author} {\bibfnamefont {K.}~\bibnamefont
  {Sch\"onhammer}},\ }\Doi {10.1088/0953-8984/21/49/495306} {\bibfield
  {journal} {\bibinfo  {journal} {J. Phys.: Condens. Matter} }\textbf
  {\bibinfo {volume} {21}},\ \bibinfo {pages} {495306} (\bibinfo {year}
  {2009})}\BibitemShut {NoStop}%
\bibitem [{\citenamefont {Levine}\ \emph {et~al.}(2012)\citenamefont {Levine},
  \citenamefont {Bantegui},\ and\ \citenamefont {Burg}}]{Levine2012}%
  \BibitemOpen
  \bibfield  {author} {\bibinfo {author} {\bibfnamefont {G.~C.}\ \bibnamefont
  {Levine}}, \bibinfo {author} {\bibfnamefont {M.~J.}\ \bibnamefont
  {Bantegui}}, \ and\ \bibinfo {author} {\bibfnamefont {J.~A.}\ \bibnamefont
  {Burg}},\ }\Doi {10.1103/PhysRevB.86.174202} {\bibfield  {journal} {\bibinfo
  {journal} {Phys. Rev. B} }\textbf {\bibinfo {volume} {86}},\ \bibinfo
  {pages} {174202} (\bibinfo {year} {2012})}\BibitemShut {NoStop}%
\bibitem [{\citenamefont {Jonckheere}\ \emph {et~al.}(2012)\citenamefont
  {Jonckheere}, \citenamefont {Stoll}, \citenamefont {Rech},\ and\
  \citenamefont {Martin}}]{Jonckheere2012}%
  \BibitemOpen
  \bibfield  {author} {\bibinfo {author} {\bibfnamefont {T.}~\bibnamefont
  {Jonckheere}}, \bibinfo {author} {\bibfnamefont {T.}~\bibnamefont {Stoll}},
  \bibinfo {author} {\bibfnamefont {J.}~\bibnamefont {Rech}}, \ and\ \bibinfo
  {author} {\bibfnamefont {T.}~\bibnamefont {Martin}},\ }\Doi
  {10.1103/PhysRevB.85.045321} {\bibfield  {journal} {\bibinfo  {journal}
  {Phys. Rev. B} }\textbf {\bibinfo {volume} {85}},\ \bibinfo {pages}
  {045321} (\bibinfo {year} {2012})}\BibitemShut {NoStop}%
\bibitem [{\citenamefont {Mah\'e}\ \emph {et~al.}(2010)\citenamefont {Mah\'e},
  \citenamefont {Parmentier}, \citenamefont {Bocquillon}, \citenamefont
  {Berroir}, \citenamefont {Glattli}, \citenamefont {Kontos}, \citenamefont
  {Pla\ifmmode~\mbox{\c{c}}\else \c{c}\fi{}ais}, \citenamefont {F\`eve},
  \citenamefont {Cavanna},\ and\ \citenamefont {Jin}}]{Mahe2010}%
  \BibitemOpen
  \bibfield  {author} {\bibinfo {author} {\bibfnamefont {A.}~\bibnamefont
  {Mah\'e}}, \bibinfo {author} {\bibfnamefont {F.~D.}\ \bibnamefont
  {Parmentier}}, \bibinfo {author} {\bibfnamefont {E.}~\bibnamefont
  {Bocquillon}}, \bibinfo {author} {\bibfnamefont {J.-M.}\ \bibnamefont
  {Berroir}}, \bibinfo {author} {\bibfnamefont {D.~C.}\ \bibnamefont
  {Glattli}}, \bibinfo {author} {\bibfnamefont {T.}~\bibnamefont {Kontos}},
  \bibinfo {author} {\bibfnamefont {B.}~\bibnamefont
  {Pla\ifmmode~\mbox{\c{c}}\else \c{c}\fi{}ais}}, \bibinfo {author}
  {\bibfnamefont {G.}~\bibnamefont {F\`eve}}, \bibinfo {author} {\bibfnamefont
  {A.}~\bibnamefont {Cavanna}}, \ and\ \bibinfo {author} {\bibfnamefont
  {Y.}~\bibnamefont {Jin}},\ }\Doi {10.1103/PhysRevB.82.201309} {\bibfield
  {journal} {\bibinfo  {journal} {Phys. Rev. B} }\textbf {\bibinfo {volume}
  {82}},\ \bibinfo {pages} {201309} (\bibinfo {year} {2010})}\BibitemShut
  {NoStop}%
\bibitem [{\citenamefont {Albert}\ \emph {et~al.}(2010)\citenamefont {Albert},
  \citenamefont {Flindt},\ and\ \citenamefont {B\"uttiker}}]{Albert2010}%
  \BibitemOpen
  \bibfield  {author} {\bibinfo {author} {\bibfnamefont {M.}~\bibnamefont
  {Albert}}, \bibinfo {author} {\bibfnamefont {C.}~\bibnamefont {Flindt}}, \
  and\ \bibinfo {author} {\bibfnamefont {M.}~\bibnamefont {B\"uttiker}},\ }\Doi
  {10.1103/PhysRevB.82.041407} {\bibfield  {journal} {\bibinfo  {journal}
  {Phys. Rev. B} }\textbf {\bibinfo {volume} {82}},\ \bibinfo {pages}
  {041407(R)} (\bibinfo {year} {2010})}\BibitemShut {NoStop}%
\bibitem [{\citenamefont {Parmentier}\ \emph {et~al.}(2012)\citenamefont
  {Parmentier}, \citenamefont {Bocquillon}, \citenamefont {Berroir},
  \citenamefont {Glattli}, \citenamefont {Pla\ifmmode~\mbox{\c{c}}\else
  \c{c}\fi{}ais}, \citenamefont {F\`eve}, \citenamefont {Albert}, \citenamefont
  {Flindt},\ and\ \citenamefont {B\"uttiker}}]{Parmentier2012}%
  \BibitemOpen
  \bibfield  {author} {\bibinfo {author} {\bibfnamefont {F.~D.}\ \bibnamefont
  {Parmentier}}, \bibinfo {author} {\bibfnamefont {E.}~\bibnamefont
  {Bocquillon}}, \bibinfo {author} {\bibfnamefont {J.-M.}\ \bibnamefont
  {Berroir}}, \bibinfo {author} {\bibfnamefont {D.~C.}\ \bibnamefont
  {Glattli}}, \bibinfo {author} {\bibfnamefont {B.}~\bibnamefont
  {Pla\ifmmode~\mbox{\c{c}}\else \c{c}\fi{}ais}}, \bibinfo {author}
  {\bibfnamefont {G.}~\bibnamefont {F\`eve}}, \bibinfo {author} {\bibfnamefont
  {M.}~\bibnamefont {Albert}}, \bibinfo {author} {\bibfnamefont
  {C.}~\bibnamefont {Flindt}}, \ and\ \bibinfo {author} {\bibfnamefont
  {M.}~\bibnamefont {B\"uttiker}},\ }\Doi {10.1103/PhysRevB.85.165438}
  {\bibfield  {journal} {\bibinfo  {journal} {Phys. Rev. B} }\textbf
  {\bibinfo {volume} {85}},\ \bibinfo {pages} {165438} (\bibinfo {year}
  {2012})}\BibitemShut {NoStop}%
\bibitem [{\citenamefont {Daley}\ \emph {et~al.}(2004)\citenamefont {Daley},
  \citenamefont {Kollath}, \citenamefont {Schollw\"{o}ck},\ and\ \citenamefont
  {Vidal}}]{Daley2004}%
  \BibitemOpen
  \bibfield  {author} {\bibinfo {author} {\bibfnamefont {A.~J.}\ \bibnamefont
  {Daley}}, \bibinfo {author} {\bibfnamefont {C.}~\bibnamefont {Kollath}},
  \bibinfo {author} {\bibfnamefont {U.}~\bibnamefont {Schollw\"{o}ck}}, \ and\
  \bibinfo {author} {\bibfnamefont {G.}~\bibnamefont {Vidal}},\ }\Doi
  {10.1088/1742-5468/2004/04/P04005} {\bibfield  {journal} {\bibinfo  {journal}
  {J. Stat. Mech.} \bibinfo {pages} {P04005}} (\bibinfo {year}
  {2004})}\BibitemShut {NoStop}%
\bibitem [{\citenamefont {Bohr}\ \emph {et~al.}(2006)\citenamefont {Bohr},
  \citenamefont {Schmitteckert},\ and\ \citenamefont {W\"{o}lfle}}]{Bohr2006}%
  \BibitemOpen
  \bibfield  {author} {\bibinfo {author} {\bibfnamefont {D.}~\bibnamefont
  {Bohr}}, \bibinfo {author} {\bibfnamefont {P.}~\bibnamefont {Schmitteckert}},
  \ and\ \bibinfo {author} {\bibfnamefont {P.}~\bibnamefont {W\"{o}lfle}},\
  }\Doi {dx.doi.org/10.1209/epl/i2005-10377-6} {\bibfield  {journal} {\bibinfo
  {journal} {Europhys. Lett.} }\textbf {\bibinfo {volume} {73}},\ \bibinfo
  {pages} {246} (\bibinfo {year} {2006})}\BibitemShut {NoStop}%
\bibitem [{\citenamefont {Carr}\ \emph {et~al.}(2011)\citenamefont {Carr},
  \citenamefont {Bagrets},\ and\ \citenamefont {Schmitteckert}}]{Carr2011}%
  \BibitemOpen
  \bibfield  {author} {\bibinfo {author} {\bibfnamefont {S.~T.}\ \bibnamefont
  {Carr}}, \bibinfo {author} {\bibfnamefont {D.~A.}\ \bibnamefont {Bagrets}}, \
  and\ \bibinfo {author} {\bibfnamefont {P.}~\bibnamefont {Schmitteckert}},\
  }\Doi {10.1103/PhysRevLett.107.206801} {\bibfield  {journal} {\bibinfo
  {journal} {Phys. Rev. Lett.} }\textbf {\bibinfo {volume} {107}},\ \bibinfo
  {pages} {206801} (\bibinfo {year} {2011})}\BibitemShut {NoStop}%
\bibitem [{\citenamefont {Chien}\ \emph {et~al.}(2011)\citenamefont {Chien},
  \citenamefont {Zwolak},\ and\ \citenamefont {Ventra}}]{Chien2012a}%
  \BibitemOpen
  \bibfield  {author} {\bibinfo {author} {\bibfnamefont {C.~C.}\ \bibnamefont
  {Chien}}, \bibinfo {author} {\bibfnamefont {M.}\ \bibnamefont {Zwolak}}, \
  and\ \bibinfo {author} {\bibfnamefont {M.}~\bibnamefont {Di Ventra}},\
  }\Doi {10.1103/PhysRevA.85.041601} {\bibfield  {journal} {\bibinfo
  {journal} {Phys. Rev. A} }\textbf {\bibinfo {volume} {85}},\ \bibinfo
  {pages} {041601(R) } (\bibinfo {year} {2012})}\BibitemShut {NoStop}%
\bibitem [{\citenamefont {Chien}\ \emph {et~al.}(2013)\citenamefont {Chien},
  \citenamefont {Gruss}, \citenamefont {Di Ventra},\ and\ \citenamefont
  {Zwolak}}]{Chien2013}%
  \BibitemOpen
  \bibfield  {author} {\bibinfo {author} {\bibfnamefont {C.-C.}~\bibnamefont
  {Chien}}, \bibinfo {author} {\bibfnamefont {D.}~\bibnamefont {Gruss}},
  \bibinfo {author} {\bibfnamefont {M.}~\bibnamefont {Di Ventra}}, \ and\ \bibinfo
  {author} {\bibfnamefont {M.}~\bibnamefont {Zwolak}},\ }\Doi
  {10.1088/1367-2630/15/6/063026
  } {\bibfield  {journal} {\bibinfo  {journal}
  {New J. Phys.} }\textbf {\bibinfo {volume} {15}},\ \bibinfo {pages}
  {063026} (\bibinfo {year} {2013})}\BibitemShut {NoStop}%
\bibitem [{\citenamefont {Brantut}\ \emph {et~al.}(2012)\citenamefont
  {Brantut}, \citenamefont {Meineke}, \citenamefont {Stadler}, \citenamefont
  {Krinner},\ and\ \citenamefont {Esslinger}}]{Brantut2012}%
  \BibitemOpen
  \bibfield  {author} {\bibinfo {author} {\bibfnamefont {J.-P.}\ \bibnamefont
  {Brantut}}, \bibinfo {author} {\bibfnamefont {J.}~\bibnamefont {Meineke}},
  \bibinfo {author} {\bibfnamefont {D.}~\bibnamefont {Stadler}}, \bibinfo
  {author} {\bibfnamefont {S.}~\bibnamefont {Krinner}}, \ and\ \bibinfo
  {author} {\bibfnamefont {T.}~\bibnamefont {Esslinger}},\ }\Doi
  {10.1126/science.1223175} {\bibfield  {journal} {\bibinfo  {journal}
  {Science} }\textbf {\bibinfo {volume} {337}},\ \bibinfo {pages} {1069}
  (\bibinfo {year} {2012})}\BibitemShut {NoStop}%
\bibitem [{\citenamefont {Krinner}\ \emph {et~al.}}(2014)]{Krinner2014}%
  \BibitemOpen
  \bibfield  {author} {\bibinfo {author} {\bibfnamefont {S.}~\bibnamefont
  {Krinner}}, {\bibfnamefont {D.}~\bibnamefont
  {Stadler}}, {\bibfnamefont {D.}~\bibnamefont
  {Husmann}}, {\bibfnamefont {J.-P}~\bibnamefont
  {Brantut}}\ and\ \bibinfo {author} {\bibfnamefont {T.}~\bibnamefont
  {Esslinger}},\ }\href@noop {} {\bibinfo  {journal}
  {arXiv:1404.6400}}\BibitemShut {NoStop}%
\bibitem [{\citenamefont {Chien}\ and\ \citenamefont
  {Di Ventra}(2012)}]{Chien2012b}%
  \BibitemOpen
  \bibfield  {author} {\bibinfo {author} {\bibfnamefont {C.-C.}~\bibnamefont
  {Chien}}\ and\ \bibinfo {author} {\bibfnamefont {M.}~\bibnamefont
  {Di Ventra}},\ }\Doi {10.1209/0295-5075/99/40003} {\bibfield
  {journal} {\bibinfo  {journal} {Europhys. Lett.} }\textbf
  {\bibinfo {volume} {99}},\ \bibinfo {pages} {40003} (\bibinfo {year}
  {2012})}\BibitemShut {NoStop}%
\bibitem [{\citenamefont {Chien}\ \emph {et~al.}}(2014)]{Chien2014}%
  \BibitemOpen
  \bibfield  {author} {\bibinfo {author} {\bibfnamefont {C.-C.}~\bibnamefont
  {Chien}}, {\bibfnamefont {M.}~\bibnamefont
  {Di Ventra}}\ and\ \bibinfo {author} {\bibfnamefont {M.}~\bibnamefont
  {Zwolak}},\ }\href@noop {} {\bibinfo  {journal}
  {arXiv:1403.0511}}\BibitemShut {NoStop}%
\bibitem [{Note1()}]{Note1}%
  \BibitemOpen
  \bibinfo {note} {The continuum description can be recovered by taking the
  limits $a\to 0,\protect \tmspace +\thickmuskip {.2777em} M\to \infty $, where
  $M$ is the number of lattice sites, keeping the length of the system $L=Ma$
  constant.}\BibitemShut {Stop}%
\bibitem [{\citenamefont {Su}\ \emph {et~al.}(1979)\citenamefont {Su},
  \citenamefont {Schrieffer},\ and\ \citenamefont {Heeger}}]{Su1979}%
  \BibitemOpen
  \bibfield  {author} {\bibinfo {author} {\bibfnamefont {W. P.}~\bibnamefont
  {Su}}, \bibinfo {author} {\bibfnamefont {J. R.}~\bibnamefont {Schrieffer}}, \
  and\ \bibinfo {author} {\bibfnamefont {A. J.}~\bibnamefont {Heeger}},\ }\Doi
  {10.1103/PhysRevLett.42.1698} {\bibfield  {journal} {\bibinfo  {journal}
  {Phys. Rev. Lett.} }\textbf {\bibinfo {volume} {42}},\ \bibinfo {pages}
  {1698} (\bibinfo {year} {1979})}\BibitemShut {NoStop}%
\bibitem [{\citenamefont {Su}\ \emph {et~al.}(1980)\citenamefont {Su},
  \citenamefont {Schrieffer},\ and\ \citenamefont {Heeger}}]{Su1980}%
  \BibitemOpen
  \bibfield  {author} {\bibinfo {author} {\bibfnamefont {W. P.}\ \bibnamefont
  {Su}}, \bibinfo {author} {\bibfnamefont {J. R.}\ \bibnamefont
  {Schrieffer}}, \ and\ \bibinfo {author} {\bibfnamefont {A. J.}\ \bibnamefont
  {Heeger}},\ }\Doi {10.1103/PhysRevB.22.2099} {\bibfield  {journal} {\bibinfo
  {journal} {Phys. Rev. B} }\textbf {\bibinfo {volume} {22}},\ \bibinfo
  {pages} {2099} (\bibinfo {year} {1980})}\BibitemShut {NoStop}%
\bibitem [{\citenamefont {Datta}(1995)}]{Datta1995}%
  \BibitemOpen
  \bibfield  {author} {\bibinfo {author} {\bibfnamefont {S.}~\bibnamefont
  {Datta}},\ }\href@noop {} {\emph {\bibinfo {title} {Electronic transport in
  mesoscopic systems}}}\ (\bibinfo  {publisher} {Cambridge University Press,
  New York},\ \bibinfo {year} {1995})\BibitemShut {NoStop}%
\bibitem [{\citenamefont {Gurvitz}\ and\ \citenamefont
  {Prager}(1996)}]{Gurvitz1996}%
  \BibitemOpen
  \bibfield  {author} {\bibinfo {author} {\bibfnamefont {S.~A.}\ \bibnamefont
  {Gurvitz}}\ and\ \bibinfo {author} {\bibfnamefont {Ya.~S.}\ \bibnamefont
  {Prager}},\ }\Doi {10.1103/PhysRevB.53.15932} {\bibfield  {journal} {\bibinfo
   {journal} {Phys. Rev. B} }\textbf {\bibinfo {volume} {53}},\ \bibinfo
  {pages} {15932} (\bibinfo {year} {1996})}\BibitemShut {NoStop}%
\bibitem [{\citenamefont {Feynman}\ \emph {et~al.}(1965)\citenamefont
  {Feynman}, \citenamefont {Leighton},\ and\ \citenamefont
  {Sands}}]{Feynman1965}%
  \BibitemOpen
  \bibfield  {author} {\bibinfo {author} {\bibfnamefont {R.~P.}\ \bibnamefont
  {Feynman}}, \bibinfo {author} {\bibfnamefont {R.~B.}\ \bibnamefont
  {Leighton}}, \ and\ \bibinfo {author} {\bibfnamefont {M.}~\bibnamefont
  {Sands}},\ }\href@noop {} {\emph {\bibinfo {title} {The Feynman lectures on
  physics}}}\ (\bibinfo  {publisher} {Addison-Wesley, Reading, Massachusetts},\
  \bibinfo {year} {1965})\BibitemShut {NoStop}%
\bibitem [{\citenamefont {Plenio}\ and\ \citenamefont
  {Knight}(1998)}]{Plenio1998}%
  \BibitemOpen
  \bibfield  {author} {\bibinfo {author} {\bibfnamefont {M.}~\bibnamefont
  {Plenio}}\ and\ \bibinfo {author} {\bibfnamefont {P.~L.}\ \bibnamefont
  {Knight}},\ }\Doi {10.1103/RevModPhys.70.101} {\bibfield  {journal} {\bibinfo
   {journal} {Rev. Mod. Phys.} }\textbf {\bibinfo {volume} {70}},\ \bibinfo
  {pages} {101} (\bibinfo {year} {1998})}\BibitemShut {NoStop}%
\bibitem [{\citenamefont {Makhlin}\ \emph {et~al.}(2001)\citenamefont
  {Makhlin}, \citenamefont {Sch\"on},\ and\ \citenamefont
  {Shnirman}}]{Makhlin2001}%
  \BibitemOpen
  \bibfield  {author} {\bibinfo {author} {\bibfnamefont {Yu.}~\bibnamefont
  {Makhlin}}, \bibinfo {author} {\bibfnamefont {G.}~\bibnamefont {Sch\"on}}, \
  and\ \bibinfo {author} {\bibfnamefont {A.}~\bibnamefont {Shnirman}},\ }\Doi
  {10.1103/RevModPhys.73.357} {\bibfield  {journal} {\bibinfo  {journal} {Rev.
  Mod. Phys.} }\textbf {\bibinfo {volume} {73}},\ \bibinfo {pages} {357}
  (\bibinfo {year} {2001})}\BibitemShut {NoStop}%
\end{thebibliography}
\end{document}